\documentclass[12pt]{article}

\usepackage{amssymb}
\usepackage{bbm}
\usepackage{epsfig}
\usepackage{dsfont}
\usepackage{amstext}
\usepackage{amsmath}
\usepackage{psfrag}
\usepackage{a4}
\usepackage{a4wide}
\usepackage{feynmp-auto}
\usepackage{amsmath, amssymb, graphics, setspace}
\usepackage{amsfonts}
\usepackage{multirow}
\usepackage{slashed}
\usepackage{amssymb}
\usepackage{array}
\usepackage{ragged2e}
\usepackage{float}
\usepackage{graphicx}
\graphicspath{ {Images/} }
\graphicspath{ {feyn/} }
\usepackage{xparse}
\usepackage{amsmath, amssymb, graphics, setspace}
\usepackage{rotating}
\usepackage{float}
\usepackage{graphicx}
\usepackage{graphics}
\usepackage{makeidx}
\RequirePackage{luatex85}
\usepackage{tikz}
\usepackage{hyperref}
\usepackage{subfig}
\usepackage[utf8]{inputenc}
\newcommand{\mathsym}[1]{{}}
\newcommand{\unicode}[1]{{}}

\title{}
\begin{document}
 	\begin{center}
	\baselineskip 20pt 
	{\Large\bf Sneutrino Tribrid Inflation, Metastable Cosmic Strings and Gravitational Waves }

	\vspace{1cm}

{\large 
	Muhammad Atif Masoud$^{a,}$\footnote{E-Mail: atifmasood23@gmail.com}, Mansoor Ur Rehman$^{a,}$\footnote{ E-mail: mansoor@qau.edu.pk},	 
and Qaisar Shafi$^{b,}$\footnote{ E-mail:  qshafi@udel.edu}  
} 
\vspace{.5cm}

{\baselineskip 20pt \it
	$^a$Department of Physics, Quaid-i-Azam University , \\ 
	Islamabad 45320, Pakistan \\
	\vspace{2mm} 
	$^b$Bartol Research Institute, Department of Physics and Astronomy, \\
	University of Delaware, Newark, DE 19716, USA \\
}

\vspace{1cm}
\end{center}

\begin{abstract}
We present a successful realization of sneutrino tribrid inflation model based on a gauged $U(1)_{B-L}$ extension of Minimal Supersymmetric Standard Model (MSSM). A single interaction term involving the $B-L$ Higgs field and the right-handed neutrinos serves multiple purposes. These include the generation of heavy Majorana masses for the right-handed neutrinos to provide an explanation for the tiny neutrino masses via the seesaw mechanism, a realistic scenario for reheating and non-thermal leptogenesis with a reheat temperature as low as $10^6$~GeV, and a successful realization of inflation with right-handed sneutrino as the inflaton. The matter parity which helps avoid rapid proton decay survives as a $Z_{2}$ subgroup of a $U(1)$ $R$-symmetry. Depending on the  choice of model parameters yields the following predicted range of the tensor to scalar ratio, $3 \times 10^{-11}\lesssim r\lesssim 7\times 10^{-4}$ ($ 6 \times 10^{-7} \lesssim r \lesssim 0.01 $), and the running of the scalar spectral index, $-0.00022 \lesssim dn_s/d\ln k \lesssim -0.0026$ ($-0.00014 \lesssim dn_s/d\ln k  \lesssim 0.005$), along with the $B-L$ breaking scale, $ 3 \times 10^{14}\lesssim M/ \text{GeV}\lesssim 5 \times 10^{15}$ ($ 6 \times 10^{15}\lesssim M/ \text{GeV}\lesssim 2 \times 10^{16}$),  calculated at the central value of the scalar spectral index, $n_s =0.966$, reported by Planck 2018. The possibility of realizing metastable cosmic strings in a grand unified theory setup is briefly discussed. The metastable cosmic string network admits string tension values in the range $10^{-8} \lesssim G\mu_s \lesssim 10^{-6}$, and predicts a stochastic gravitational wave background lying within the 2-$\sigma$ bounds of the recent NANOGrav 12.5-yr data. 
\end{abstract}

\section{Introduction}
Supersymmetric (SUSY) hybrid inflation \cite{Dvali:1994ms,Copeland:1994vg,Senoguz:2004vu,Rehman:2009nq,Rehman:2009yj,Linde:1997sj} offers a natural setup for linking inflation with grand unified theory (GUT) based particle physics models. This connection is further strengthened in the tribrid inflation framework \cite{Antusch:2004hd,Antusch:2008pn,Antusch:2009ef,Antusch:2010mv,Antusch:2009vg,Antusch:2015tha}, an interesting extension of supersymmetric hybrid inflation,
where a matter field can be employed to realize inflation. One of the simplest candidates
for tribrid inflation is the sneutrino, the superpartner of the right-handed
neutrino. An early model of sneutrino tribrid inflation was introduced in \cite{Antusch:2004hd}
where the sneutrino field was taken to be a gauge singlet. However, a $Z_4$ symmetry is introduced to restrict the structure of the superpotential, which could lead a domain wall problem from $Z_4$ breaking at the end of inflation. To eliminate this problem  higher order $Z_4$ symmetry breaking terms are introduced. A pseudosmooth version of tribrid inflation \cite{Antusch:2012bp} can be employed to avoid the monopole problem, if present. An $SU(5)$  example is discussed in \cite{Masoud:2019gxx}. For a comprehensive discussion of tribrid inflation with gauge non-singlet matter field inflaton, see \cite{Antusch:2010va}.
For chaotic sneutrino inflation models, see \cite{Murayama:1992ua,Murayama:1993xu,Nakayama:2013nya,Murayama:2014saa}. Also, see \cite{Kallosh:2016sej} for sneutrino inflation with $\alpha$-attractors. An interesting and novel possibility in which the inflaton emerges as a superposition of the Higgs, squark and slepton is given in \cite{Tavartkiladze:2019svd}.

In this paper a gauge non-singlet sneutrino tribrid inflation model is constructed within a $U(1)_{B-L}$ extension of minimal supersymmetric standard model (MSSM), where $B$ and $L$ denote the baryon and lepton numbers respectively. This realization is based on the standard version of tribrid model first discussed in \cite{Antusch:2004hd} where inflation ends by a waterfall phase. However, an additional vectorlike neutrino superfield is required to yield a suitable D-flat direction for inflation. This addition is an essential feature of realizing inflation with a gauge non-singlet matter field \cite{Antusch:2010va}. In the superpotential, a leading order non-renormalizable term involving the $B-L$ Higgs and neutrino superfields plays a crucial role not only in realizing sneutrino tribrid inflation but also in providing the intermediate masses for right-handed neutrinos, which are necessary to explain the observed tiny neutrino masses via the seesaw mechanism. Furthermore, the same term is essential for reheating and leptogenesis. In particular, non-thermal leptogenesis yields a reheat temperature as low as $T_r =10^6$ GeV. This can avoid the gravitino problem, usually encountered in supergravity models of inflation \cite{Kawasaki:2004qu,Kawasaki:2008qe,Kawasaki:2017bqm,Khlopov:1984pf}, for a somewhat wider range of the gravitino mass. The same term if generated at a renormalizable level does not lead to tribrid inflation. With renormalizable terms in a similar model setup, chaotic inflation driven by a quartic potential associated with the $B-L$ Higgs field is discussed in \cite{Pallis:2017xfo}. Also see \cite{Senoguz:2005bc}, for SUSY hybrid inflation  in a $U(1)_{B-L}$ extension of MSSM.

The one loop radiative corrections and the supergravity corrections arising from the non-minimal K\"ahler potential make equally important contribution for the realization of successful inflation in agreement with the latest Planck 2018 data \cite{Aghanim:2018eyx,Akrami:2018odb}. The soft SUSY breaking terms, on the other hand, have a negligible effect on  inflationary predictions. This is in contrast to hybrid inflation models where the contributions from the soft SUSY breaking terms are crucial for the model prediction for the scalar spectral index to be consistent with the latest experimental data in a minimal canonical K\"ahler potential setup \cite{Rehman:2009nq,Rehman:2009yj}. The relative importance of various terms in a conventional model of tribrid inflation is discussed in \cite{Antusch:2012jc}.

The breaking of $U(1)_{B-L}$ gives rise to stable cosmic string network and can put stringent bounds on the model parameter space. However, if these cosmic strings are metastable then the bounds can be relaxed. We briefly discuss such a realization in a GUT setup based on $SO(10)$. As discussed in \cite{Buchmuller:2019gfy} this type of embedding leads to the production of a metastable cosmic sting network which can decay via the Schwinger production of monopole-antimonopole pairs. This decay generates a stochastic gravitational wave background which is in the range of ongoing and future gravitational wave (GW) experiments. We compare our model predictions with the recent bounds from the NANOGrav 12.5-yr data \cite{Arzoumanian:2020vkk}. We also highlight a parameter space for the realization of observable primordial gravitational waves from inflation.
\section{Superpotential with $U(1)_{B-L}$ Symmetry}
The superpotential for the realization of tribrid inflation in a $U(1)_{B-L} $ extension of MSSM can be written as
\begin{eqnarray}\label{sup}
W &=& \kappa S(\Phi \overline{\Phi}  - M^{2}) +\lambda S H_{u} H_{d} \nonumber\\
  &+& y_{ij}^{(U)}Q_{i}U_{j}^{c}H_{u}+y_{ij}^{(D)}Q_{i}D_{j}^{c}H_{d}+y_{ij}^{(L)}L_{i}E_{j}^{c}H_{d}+y_{ij}^{(\nu)}L_{i}H_{u}N_{j}^{c}\nonumber \\
  &+& \frac{\lambda_{ij}}{M_{*}}(\overline{\Phi})^{2}N_{i}^{c}N_{j}^{c}  +\frac{\tilde{\lambda}}{M_{*}}(\Phi)^{2}\mathcal{\overline N}^{c}\mathcal{\overline N}^{c},
 \end{eqnarray}
where $\kappa, \lambda, \lambda_{i}, \lambda_{ij}$ and $\tilde{\lambda} $ are dimensionless couplings,  $y_{ij}^{(U)}, y_{ij}^{(D)}, y_{ij}^{(L)}$ and $y_{ij}^{(\nu)}$ are the Yukawa couplings, $N_{i}^{c}=(N_{1}^{c},N_{2}^{c},N_{3}^{c},\mathcal{N}^{c})$, and $M_*$ is some superheavy mass.  In addition to the local symmetry, the superpotential in Eq.~(\ref{sup}) possesses three global symmetries, namely, $ U(1)_{R}, U(1)_{B}$ and $U(1)_{L}$ with $R(W) = 2$, $B(W) = 0$ and $L(W) = 0$. The charge assignments under these symmetries of the various matter and Higgs superfields are given in  Table.~\ref{table:1}. To simplify the discussion we also assume an extra $Z_{2}$  symmetry under which only the $\mathcal{\overline N}^{c}$ superfield is odd. So the terms linear in $\mathcal{\overline N}^{c}$ are forbidden in $W$.

\begin{table}[th]
 	{\centering
 		{
 			\begin{tabular}{||c||c||c||c||c||}
 				\hline	
 				{\bf 	Superfields} &  ${\bf R}$ & $  {\bf B}$ & $  {\bf L}$ & $ {\bf  B-L}$     \\
 				\hline \hline
 				$E^c_i$ & $1$ & $ 0$ & $ -1$ & $ 1$     \\ 
 				\hline 
 				$N^c_i $ & $1$ & $ 0$ & $ -1$ & $ 1$     \\ 
 				\hline 
 				$ \mathcal{\overline N}^{c}$ & $1$ & $ 0$ & $ 1$ & $ -1$     \\ 
 				\hline 
 				$L_i $ & $1$ & $ 0$ & $ 1$ & $ -1$     \\ 
 				\hline 
 				$U^c_i $ & $1$ & $-1/3$ & $ 0$ & $ 1/3$     \\ 
 				\hline 
 				$D^c_i$ & $1$ & $ -1/3$ & $ 0$ & $ 1/3$     \\ 
 				\hline 
 				$Q_i$ & $1$ & $ 1/3 $ & $ 0$ & $ 1/3$     \\ 
 				\hline 
 				$H_{d}$ & $0$ & $ 0$ & $ 0$ & $ 0$     \\ 
 				\hline 
 				$H_{u}$ & $0$ & $ 0$ & $0$ & $ 0$     \\ 
 				\hline 
 				$S$ & $2$ & $ 0$ & $ 0$ & $ 0$     \\ 
 				\hline 
 				$\Phi$ & $0$ & $ 0 $ & $ -1$ & $ 1$     \\ 
 				\hline 
 				$\overline{\Phi}$ & $0$ & $ 0$ & $ 1$ & $ -1$     \\ 
 				\hline
 			\end{tabular} 
 			\caption{The global and local charges of superfields in present model.} \label{table:1}
 		}
 		\centering}
 \end{table}

The gauge singlet superfield, $S$, is required by the $R$-symmetric tribrid framework for providing a flat direction for realizing inflation with a natural setup for spontaneous breaking of the underlying gauge symmetry. Note that the scalar component of $S$ plays the role of inflaton in hybrid inflation, whereas here in tribrid inflation it remains stabilized at its minimum during inflation as described below. The scalar components of the conjugate pair of Higgs superfields $\Phi$ and $\overline{\Phi}$ break the $U(1)_{B-L}$ gauge symmetry by attaining a VEV equal to the gauge symmetry breaking scale $M$. 
The second term in the first line of the superpotential describes the MSSM $\mu$-term with $\mu =\kappa \langle S \rangle \sim(\lambda/\kappa)m_{3/2}$. This solves the MSSM $\mu$ problem once the $S$ field acquires a non-zero VEV equal to $\kappa\,m_{3/2}$  via the soft SUSY breaking terms in a supergravity framework \cite{Dvali:1997uq}. This term identifies an important class of hybrid inflation models known as  $\mu$-hybrid inflation recently considered in \cite{Okada:2015vka,Rehman:2017gkm,Lazarides:2020zof}.

The global SUSY minimum  occurs at
\begin{eqnarray}
\left< \Phi \overline{\Phi} \right> =  M^2, \quad \left< S \right> = 0, \quad\left< N_{i}^{c} \right> = 0,\quad\left< \mathcal{\overline N}^{c}\right> = 0,\quad\left< H_{u} \right> = 0,\quad\left< H_{d} \right> = 0.
\end{eqnarray}
After the breaking of $U(1)_{B-L}$ gauge symmetry, the last two terms in Eq.~(\ref{sup}) give rise to Majorana mass terms,  $ M_{ij}^RN_{i}^{c}N_{j}^{c} +  M^R \mathcal{\overline N}^{c}\mathcal{\overline N}^{c}$, for the right-handed neutrinos with 
\begin{eqnarray}\label{majmass}
               M_{ij}^R = \lambda_{ij}\left(\frac{M }{M_{*}}\right)M, \quad  M^R= \tilde{\lambda}\left(\frac{M }{M_{*}}\right)M.
\end{eqnarray}
With natural values of the couplings $(\lambda_{ij},\,\tilde{\lambda})\sim \mathcal{O} (1)$ and the superheavy scale $M_*  \gtrsim 10\,M$,  we obtain Majorana masses $ \lesssim 10^{14}$~GeV with the gauge symmetry breaking scale $M \sim 10^{15}$~GeV. Therefore, the light neutrino masses are naturally generated via the seesaw mechanism. 
   
  The $R$-parity that is usually employed to prevent rapid proton decay mediated by the dimension four operators appears as a $Z_{2}$ subgroup of $U(1)_{R}$ symmetry in the present model. However, proton is essentially stable due to other global and local symmetries described in Table.~\ref{table:1}. The domain wall problem associated with $Z_{2}$ $R$-parity is avoided because this symmetry survives after the spontaneous breaking of $U(1)_R$ symmetry. Lastly, with $R$-parity we obtain a stable lightest SUSY particle (LSP) as a plausible cold dark matter candidate.  
\section{Global Supersymmetric Potential}
  The superpotential terms relevant for sneutrino tribrid inflation are
  \begin{eqnarray}
 W  \supset \kappa S(\Phi \overline{\Phi} - M^{2}) +\frac{\lambda_{ij}}{M_{*}}(\overline{\Phi})^{2}N^{c}_i N^{c}_j +\frac{\tilde{\lambda}}{M_{*}}(\Phi)^{2}\mathcal{\overline N}^{c}\mathcal{\overline N}^{c}.
 \label{winf}
\end{eqnarray}
The global SUSY $F$-term scalar potential obtained from the above superpotential is given by
\begin{eqnarray}%
V_F  & = & \left| \frac{\partial W}{\partial z_i} \right|^2  \supset \kappa^2 \left| \Phi \overline{\Phi} - M^{2} \right|^2 + \left|\frac{(\lambda_{ij} +\lambda_{ji})\overline{\Phi}^{2} N_{i}^c}{M_{*}}\right|^{2} +  \left|\frac{\tilde{2 \lambda} \Phi^{2} \mathcal{\overline N}^{c}}{M_{*}}\right|^{2}    \nonumber \\
  &+& \left|\kappa S \overline{\Phi} + \frac{2 \tilde{\lambda} \mathcal{\overline N}^{c}\mathcal{\overline N}^{c} \Phi }{M_{*}} \right|^{2} + \left|\kappa S \Phi+\frac{2\lambda_{ij} N_{i}^c N_{j}^c \overline{\Phi} }{M_{*}}\right|^{2},
\end{eqnarray}%
where $z_{i} \in ( S,\Phi,\overline{\Phi},\mathcal{\overline N}^{c},N^c )$. With $\Phi^* = \overline{\Phi}$ the relevant part of the global SUSY D-term scalar potential reads
\begin{eqnarray}
 &V_{D} \propto g^{2}_{B-L}(|N_{1}^{c}|^{2}+|N_{2}^{c}|^{2}+|N_{3}^{c}|^{2}+|\mathcal{ N}^{c} |^{2}-|\mathcal{\overline N}^{c}|^{2})^2,
\end{eqnarray}
assuming all other fields are stabilized in their respective global SUSY minimum. Here, $g_{B-L}$ is the gauge coupling of $U(1)_{B-L}$. In general, $F$-term inflation can be realized with all four $N_i^c$ fields aligned along $\mathcal{\overline N}^{c}$ field to achieve D-flat direction. Here, we consider the D-flat direction $N \equiv |{ N_{1}}^{c} |=|\mathcal{\overline N}^{c}|$ with $N_{i\neq 1}^{c}=0$ for the realization of sneutrino inflation. The inclusion of at least one conjugate pair $\mathcal{\overline N}^{c},\,\mathcal{N}^{c}$ is therefore crucial for the realization of such a D-flat inflationary direction.

The three-field scalar potential $V_3 = V_F + V_D$ takes the following form in the D-flat direction ($|\overline{\Phi}|=|\Phi|$),
\begin{eqnarray}%
V_3(x,y,z)  &=&  \kappa^2 M^{4}\Bigg((z^2 -1)^2 + \left( x z + \frac{2 \tilde{\lambda}}{\kappa}  \left(\frac{M}{M_*}\right) y^{2}z\right)^{2}  + \left( x z+\frac{2\lambda_{ij}}{\kappa}  \left(\frac{M}{M_*}\right) y^{2} z\right)^{2}  \nonumber\\&&
+ \left(\frac{(\lambda_{ij}+\lambda_{ji})}{\kappa}\left(\frac{M}{M_*}\right) z^{2}  y \right)^{2} +  \left(\frac{2\tilde{ \lambda}}{\kappa} \left(\frac{M}{M_*}\right) z^{2}  y \right)^{2}\Bigg),
\end{eqnarray}%
where the normalized fields $x$, $y$ and $z$ are defined as
\begin{eqnarray}\label{dimless}
x =\frac{|S|}{M},\quad y=\frac{|N|}{M},\quad z=\frac{|\Phi|}{M}.
\end{eqnarray}
To obtain the above potential we assume that the phases of the fields involved have been stabilized before the onset of observable inflation. 
\section{Inflationary Potential with Supergravity Corrections}
\begin{figure}[t]
	\centering
	\includegraphics[width=10cm]{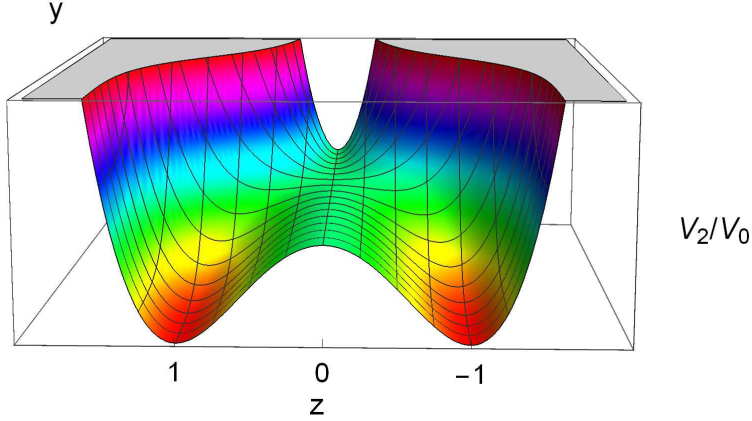}
	\caption{The normalized two-field potential $V_{2}/V_0$ as a function of $z = \frac{|\Phi|}{M}$ and $y = \frac{|N|}{M}$ with $S=0$, $\gamma = \delta' = 0$, $y_c=1$ and $V_0 = \kappa^2 M^4$.}  \label{3Dplot}
\end{figure}

The supergravity (SUGRA) corrections in the $F$-term scalar potential can be obtained from the following expression, 
\begin{eqnarray}
V_F = e^{K/m_{P}^{2}}\left(K_{ij}^{-1}D_{z_{i}}WD_{z_{j}^{*}}W^{*} -3m_{P}^{-2}|W|^{2}\right),
\end{eqnarray}
where
\begin{eqnarray}
D_{z_{i}}W = \frac{\partial W}{\partial z_{i}}+\frac{1}{m_{P}^{2}}\frac{\partial K}{\partial z_{i}}W, \,\,
 K_{ij} = \frac{\partial^{2}K}{\partial z_{i}\partial z_{j}^{*}}, \,\,
D_{z_{j}^{*}}W^{*} = (D_{z_{i}}W)^{*},
\end{eqnarray}
and $m_P = 2.43 \times 10^{18}$~GeV is the reduced Planck mass. For the K\"ahler potential $K$ we consider the following relevant terms,
\begin{eqnarray}%
K &=& |S|^{2}+|\Phi|^{2}+|\overline{\Phi}|^{2}+|N_1|^{2}+|\mathcal{\overline N}^{c}|^{2} \nonumber\\
&+&\kappa_{S}\frac{|S|^{4}}{4m_{P}^{2}}+\kappa_{N} \frac{|N_1|^{4}}{4 m_{P}^{2}} +\kappa_{\mathcal{\overline N}^{c}} \frac{|\mathcal{\overline N}^{c}|^{4}}{4 m_{P}^{2}}\nonumber \\
&+&\kappa_{SN} \frac{|S|^{2}|N_1|^{2}}{m_{P}^{2}} +\kappa_{S\mathcal{\overline N}^{c}} \frac{|S|^{2}|\mathcal{\overline N}^{c}|^{2}}{m_{P}^{2}}
+ \kappa_{N \mathcal{\overline N}^{c}} \frac{|N_1|^{2}|\mathcal{\overline N}^{c}|^{2}}{m_{P}^{2}}. 
\end{eqnarray}%
The $\kappa_S$-term in $K$ generates a Hubble mass, $\sqrt{-3\,\kappa_{S}} H$, for the $S$ field for $\kappa _S \lesssim-\frac{1}{3}$ with $H^{2} \simeq \frac{\kappa^2 M ^4}{3 m_P^2}$. The $S$ field, thus, quickly stabilizes in its global minimum $\left< S \right> = 0$. Including SUGRA corrections, the two-field scalar potential $V_2$ with $S=0$ can now be written in the D-flat direction as
\begin{eqnarray}\label{V2POT}
V_2(y,z) &=& \kappa^2 M^4 \Bigg((z^{2}-1)^{2} +  \frac{4(\lambda_{11}^{2}+\tilde{\lambda}^{2})}{\kappa^2} \left(\frac{M}{M_{*}}\right)^{2}\left(z^{4} y^{2}+z^{2} y^{4}\right)
 + \gamma\left(\frac{M}{m_{P}}\right) ^{2} y^{2} \nonumber  \\
 &+& \,\delta\left(\frac{M}{m_{P}}\right) ^{4} y^{4}+\cdots. \Bigg),
\end{eqnarray}
where,
\begin{eqnarray}\label{gmdl}
\gamma &=& 2 -\kappa_{SN} -\kappa_{S\mathcal{\overline N}^{c}}, \nonumber \\
\delta' &=& 2 +\frac{1}{4}\kappa_{N} +\frac{1}{4}\kappa_{\mathcal{\overline N}^{c}}+ \kappa_{N\mathcal{\overline N}^{c}} -2\kappa_{SN}+\kappa_{SN}^{2}  -2\kappa_{S\mathcal{\overline N}^{c}} + 2\kappa_{SN}\kappa_{S\mathcal{\overline N}^{c}} + \kappa_{S\mathcal{\overline N}^{c}}^2.
\end{eqnarray}
The above two-field scalar potential with $\gamma = \delta' = 0$ is displayed in Fig.~\ref{3Dplot} as a function of $z$ and $y$. In this figure a flat trajectory suitable for sneutrino inflation and a waterfall region to end inflation are clearly visible.     

The mass squared of waterfall Higgs field along the track  $\Phi=0$ is given by
\begin{eqnarray}%
m_{\Phi}^{2} = 8 M^{2}\left(\frac{M}{M_{*}}\right)^{2} (\tilde{\lambda}^2+\lambda_{11}^2)y^4 -  4 \kappa^{2} M^{2}.
\end{eqnarray}
Inflation occurs in the valley with $m_{\Phi}^{2}$ positive and ends when $m_{\Phi}^{2}$ becomes negative. This happens when the $y$ field value drops below its critical value $y_c$ defined as 
\begin{eqnarray}\label{waterfall}
y_{c} = \frac{\sqrt{\kappa }}{ 2^{1/4}  (\tilde{\lambda}^2+\lambda_{11}^2)^{1/4}}\left(\frac{ M_{*}}{M}\right)^{1/2} = \frac{\sqrt{\kappa }}{ \sqrt{2\,\lambda}}\left(\frac{ M_{*}}{M}\right)^{1/2},
\end{eqnarray}
where for simplicity we set $\lambda = \tilde{\lambda}=\lambda_{11}$. After the waterfall point ($y=y_c$) the system rapidly settles in its global minimum and the $B-L$ gauge symmetry is spontaneously broken. 
During the inflationary trajectory ($\Phi = 0$), as shown in Fig.~\ref{3Dplot}, we obtain the following effective single-field scalar potential:
\begin{eqnarray}
V_1 = \kappa^{2} M^{4}\left(1+\gamma \left(\frac{M}{m_{P}}\right)^{2}y^{2} + \delta'\left(\frac{M}{m_{P}}\right)^{4}y^{4}\right).  \label{V1}
\end{eqnarray}%
As we have included non-minimal terms in the K\"ahler potential, we need to redefine the relevant fields such that $K_{ij} = \delta_{ij}$. To obtain a canonically normalized inflaton field we assume $\kappa_{\mathcal{N}\mathcal{\overline{N}}^c}=0$ in order to make $K_{ij}$ diagonal during inflation. This leads to the following diagonal elements of $K_{ij}$ for the relevant neutrino fields, ($N_1,\,\mathcal{\overline{N}}^c)$,  
\begin{eqnarray}
K_{N_1 N_1^*} = 1 + \kappa_{N} \frac{|N_1|^2}{m_P^2}, \, K_{\mathcal{\overline{N}}^c  {\mathcal{\overline{N}}^c}^*} = 1 + \kappa_{\mathcal{\overline{N}}^c} \frac{|\mathcal{\overline{N}}^c|^2}{m_P^2},
\end{eqnarray}% 
with $S=0=\Phi = \overline{\Phi}$. To obtain $K_{\widehat{N}_1 \widehat{N}_1^*}=K_{\widehat{\mathcal{\overline{N}}^c}  {\widehat{\mathcal{\overline{N}}^c}^*}}= 1$ we consider the following redefinition of the neutrino fields, 
\begin{eqnarray}
\widehat{N}_1 = N_1 \left( 1 + \frac{\kappa_N}{2} \frac{|N_1|^2}{m_P^2} \right),  \quad \widehat{\mathcal{\overline{N}}^c} = \mathcal{\overline{N}}^c \left( 1 + \frac{\kappa_{\mathcal{\overline{N}}^c}}{2} \frac{|\mathcal{\overline{N}}^c|^2}{m_P^2} \right).
\end{eqnarray}
In the D-flat direction, $|N_1|=|\mathcal{\overline{N}}^c| =|N|$, with $\frac{(\kappa_N,\, \kappa_{\mathcal{\overline{N}}^c})}{2 } \frac{|N|^2}{m_P^2} <1 $ we can obtain the following series solution for $(N_1,\,\mathcal{\overline{N}}^c)$,
\begin{eqnarray}
(N_1,\,\mathcal{\overline{N}}^c) = \widehat{N} \left( 1 - \frac{\kappa_{(N_1,\,\mathcal{\overline{N}}^c)}}{2} \left( \frac{|\widehat{N}|}{m_P} \right)^2 + \frac{3\kappa_{(N_1,\,\mathcal{\overline{N}}^c)}^2}{4} \left( \frac{|\widehat{N}|}{m_P} \right)^4 + \cdots \right). 
\end{eqnarray}
This amounts to the following modification in the scalar potential:
\begin{eqnarray}
V_1 = \kappa^{2} M^{4}\left(1+\gamma \left(\frac{M}{m_{P}}\right)^{2}y^{2} + \delta \left(\frac{M}{m_{P}}\right)^{4}y^{4}\right), 
\end{eqnarray}% 
where the redefined field, $\widehat{N}$, is represented with the old notation, $y=|\widehat{N}|/M$. One can see that only the quartic term with $\delta$ coupling is modified to,
\begin{eqnarray}
\delta= \delta' - (\kappa_N +\kappa_{\mathcal{\overline{N}}^c})+ \kappa_{N}\kappa_{SN} +\kappa_{\mathcal{\overline{N}}^c} \kappa_{S\mathcal{\overline{N}}^c},
\end{eqnarray}
after redefinition of the inflaton field. Assuming $0 \leq  |\kappa_i| \lesssim \mathcal{O}(1)$, for all coefficients $\kappa_i$ of higher-dimensional operators, we can treat $\gamma$ and $\delta$ couplings to be independent in the ranges $-2 \lesssim \gamma \lesssim 6$ and $-2 \lesssim \delta \lesssim 30$. Another important contribution to this potential comes from the radiative corrections as described below.
\section{Radiative Corrections}
The one-loop radiative corrections to scalar potential can be obtained from the following Coleman-Weinberg formula 
\begin{eqnarray}
V_{1 loop}(N) = \frac{1}{64 \pi^{2}} \text{Str}\left[\mathcal{M}^{4}(N)\left(\ln\left( \frac{\mathcal{M}^{2}(N)}{Q^{2}}\right)-\frac{3}{2} \right)\right],  \label{CWP}
\end{eqnarray}
where $\mathcal{M}$ denotes the mass matrix and $Q$ is a renormalization scale. The supertrace (Str) represents the sum over all fermionic and
bosonic degrees of freedom. In order to obtain the mass matrix $\mathcal{M}$ we calculate the masses of the relevant fields during inflation. The squares of fermionic mass $m_F$ and bosonic mass $m_B$, with $S=0,\,|\Phi|=|\overline{\Phi}|=0 $, are given by 
\begin{equation}
m_{B}^{2} = \frac{2 N ^{4}(\lambda_{11}^{2}+ \tilde{\lambda}^{2})}{M_{*}^{2}} \pm \kappa^{2} M^{2}, \quad   m_{F}^{2} = \frac{2 N ^{4}(\lambda_{11}^{2}+ \tilde{\lambda}^{2})}{M_{*}^{2}}.
\end{equation}
In calculating the above mass spectrum we have ignored the SUGRA corrections as their appearance via loop corrections is expected to be suppressed \cite{Antusch:2010mv}. The $N$-dependent contribution from the $y_{ij}^{\nu}$-Yukawa coupling is also negligible in this approximation. Moreover, the mass spectrum contribution from the gauge sector includes one real scalar, one Dirac fermion, and one gauge boson each of squared-mass equal to $2 g_{B-L}^2 |N|^2$. The supertrace over this mass spectrum vanishes and the spectrum also does not contribute to the above Coleman–Weinberg potential. Finally, including radiative 1-loop corrections (Eq.~(\ref{CWP})) along with the leading order SUGRA correction (Eq.~(\ref{V1})), the effective single-field scalar potential takes the following form:
\begin{eqnarray} \label{potential}
V  \simeq \kappa^2 M^4 \left(1 + \frac{\kappa^2}{8 \pi^2} F(w) +  \gamma \left(\frac{M}{m_{P}}\right)^{2}y^{2} + \delta\left(\frac{M}{m_{P}}\right)^{4}y^{4}\right).
\end{eqnarray}
The radiative correction is described by the function, $F(w)$, which is defined as
\begin{eqnarray}
F(w) = \frac{1}{4}[(w^4+1)\ln\big(\frac{w^4-1}{w^4}\Big)+2w^2\ln\Big(\frac{w^2+1}{w^2-1}\Big)+2\ln\Big(\frac{\kappa^2M^2w^2}{Q^2}\Big)-3],
\end{eqnarray}
where, 
\begin{eqnarray}
w \equiv \sqrt{\frac{2 N^{4}((\tilde{\lambda}^2+\lambda_{11}^2)}{M_{*}^{2}\kappa^{2}M^{2}}}= \left(\frac{y}{y_c}\right)^{2}. 
\end{eqnarray}
Note that in the present model the radiative corrections are found to play an equally important role in making inflationary predictions along with SUGRA corrections. This is generally expected for a conventional tribrid inflation model as discussed in \cite{Antusch:2012jc}.

The contribution of the soft SUSY breaking terms with TeV scale masses are usually expected to be suppressed in tribrid inflation models as described in \cite{Antusch:2012jc}. As the superpotential $W$  in Eq.~(\ref{winf}) remains zero during the inflationary period, the soft SUSY breaking $A$-term is negligible. Moreover, the soft mass term   $m_{\text{soft}}^2 |N|^2$ can be ignored as compared to the quadratic term in the above potential for $m_{\text{soft}} \ll \sqrt{\gamma} \kappa (M^2/m_P) $. This turns out to be  $\sqrt{\gamma}\,\kappa \gg 10^{-9}$ for $M \sim 10^{15}$ GeV and TeV scale soft masses. This approximation holds true for our numerical results presented in the later sections. 

It is important to note that we assume $R$ symmetry to be broken in the hidden sector such that the cosmological constant in the visible sector has the desired value. This breaking, mediated via gravity, appears in the visible sector in the form of soft SUSY breaking terms. See, for example, Sec.~6.3 of \cite{Nilles:1983ge}. As discussed in the previous paragraph, the soft SUSY breaking terms with TeV scale masses make a negligible contribution in inflationary predictions. However, after inflation, the gauge singlet field $S$ acquires a non-zero VEV with the help of these terms, which solves the $\mu$ problem in the MSSM.

Regarding the two-loop Dvali problem \cite{Dvali:1995fb} the analysis of  \cite{Antusch:2010va}, for a general class of gauge non-singlet inflation models, is also valid for our model. Using their expressions derived for the 2-loop diagrams, we obtain the following contributions to the effective squared-mass of inflaton, $\delta m^2$,
 in terms of $\mathcal{H}^2 \simeq \kappa^2 M^4/3 m_P^2$,
\begin{eqnarray}
\frac{{\delta m}^2}{\mathcal{H}^2}&\sim& \frac{\kappa^2}{(4\pi)^4 }\frac{3}{4} \left( \frac{m_P}{M}\right)^2 \left( \frac{M_R}{M}\right)^2, \\
\frac{{\delta m}^2}{\mathcal{H}^2}&\sim& \frac{\kappa^2 g^2}{(4\pi)^4 }\frac{3}{2 y^2} \left( \frac{m_P}{M}\right)^2,  \\
\frac{{\delta m}^2}{\mathcal{H}^2}&\sim& \frac{\kappa^2 g}{(4\pi)^4 }\frac{3}{2 \sqrt{2} y} \left( \frac{m_P}{M}\right)^2
\left( \frac{M_R}{M}\right). 
\end{eqnarray}
These ratios are found to be negligible in our model even for $y=1$. Thus there is no need to worry about the two-loop contributions in our model.
\section{Inflationary Slow-roll Parameters}
The leading order slow-roll parameters, $\epsilon,\,\eta$ and the next to leading order slow-roll parameter $\xi^{2}$ are given by 
\begin{eqnarray}\label{slowroll}
\epsilon(y) =\frac{1}{4}\left(\frac{m_{P}}{M}\right)^{2}\left(\frac{\partial_{y}V}{V}\right)^{2}, \, \eta(y)=\frac{1}{2}\left(\frac{m_{P}}{M}\right)^{2}\left(\frac{\partial_{y}^{2}V}{V}\right), \, \xi^{2}(y)= \frac{1}{4}\left(\frac{m_{P}}{M}\right)^{4}\left(\frac{\partial_{y}V \partial_{y}^{3}V}{V^{2}}\right), 
\end{eqnarray}
where $\partial_y$  denotes the derivative with respect to $y$. In the slow-roll approximation, $(\epsilon , \, \eta, \, \xi^{2}) \ll 1$, the scalar spectral index $n_{s}$, the tensor-to-scalar ratio $r$ and the running of the scalar spectral index $dn_{s}/d \ln k$ are given by
\begin{eqnarray}
n_{s} &\simeq & 1 + 2\eta (y_0)- 6 \epsilon (y_0), \quad r \simeq  16\epsilon (y_0), \\
\frac{dn_{s}}{d\ln k} &\simeq & 16 \epsilon (y_0) \eta (y_0)- 24 \epsilon^{2}(y_0) - 2 \xi^{2}(y_0),
\end{eqnarray} \label{eq:28}
where $y_0$ is the field value at the pivot scale which is taken to be at $k_{0} = 0.05\text{Mpc}^{-1}$. The amplitude of the scalar perturbation, $A_{s}$, is given by
\begin{eqnarray}\label{As}
A_{s}(k_{0})=\frac{1}{24\pi^{2} \epsilon(y)}\left. \left(\frac{V(y)}{m_{P}^{4}}\right)\right|_{y=y_{0}},
\end{eqnarray}
where $ A_{s}(k_{0}) = 2.142 \times 10^{-9}$ is the Planck normalization at $k_{0} = 0.05\text{Mpc}^{-1}$  \cite{Aghanim:2018eyx,Akrami:2018odb}. This constraint allows us to express $M$ in terms of $r$ as
\begin{eqnarray} \label{Mkr}
M  \simeq \left(\frac{3 A_{s}(k_{0})\pi^{2}r}{2\kappa^2}\right)^\frac{1}{4}m_{P}. 
\end{eqnarray}
The number of e-folds, $\Delta N$, from the pivot scale to the end of inflation is given by
\begin{eqnarray}\label{deltaN}
\Delta N = 2\left(\frac{M}{m_{P}}\right)^{2} \int_{y_{e}}^{y_{0}}\frac{V}{\partial_{y}V}dy, 
\end{eqnarray}
where field value at the end of inflation is $y_e = y_c$. Assuming a standard thermal history of the universe, we can express the number of e-folds, $\Delta N$, in terms of the reheat temperature, $T_{r}$, as \cite{Kolb:1990vq}
\begin{align}\label{nne}
\Delta N \simeq 50.2 +\frac{1}{3} \ln \left(\frac{T_{r}}{10^{6}\text{ GeV}}\right)+\frac{2}{3} \ln
\left(\frac{\sqrt{\kappa} M }{10^{15}\text{ GeV}}\right).
\end{align}
In estimating the numerical predictions for the various inflationary parameters we set $T_{r} = 10^6$ GeV. Such a low value for the reheat temperature also avoids the gravitino problem for a comparatively larger range of the gravitino mass \cite{Kawasaki:2004qu,Kawasaki:2008qe,Kawasaki:2017bqm,Khlopov:1984pf}. One possible realization of such a low reheat temperature and related non-thermal leptogenesis is discussed later.
\section{Results and Discussion}
The numerical predictions of the various inflationary parameters are calculated in the slow-roll approximation described above. The scalar spectral index is fixed at its central value, $n_s = 0.966$,  and we set $y_c = 1$ along with $T_r = 10^{6}$~GeV. 
After applying all these constraints we are left with only two independent parameters which can be chosen as per convenience. We scan the parametric space for successful realization of inflation by considering the two choices $\gamma < 0$, $\delta > 0$ and  $\gamma > 0$, $\delta < 0$,  both with $\kappa \lesssim 1$. As mentioned earlier, we scan the ranges $-2 \lesssim \gamma \lesssim 6$ and $-2 \lesssim \delta \lesssim 30$ for realistic inflationary solutions. 
\subsection{The Case with $\gamma < 0$, $\delta > 0$}
\begin{figure}[ht]
\includegraphics[width=8cm]{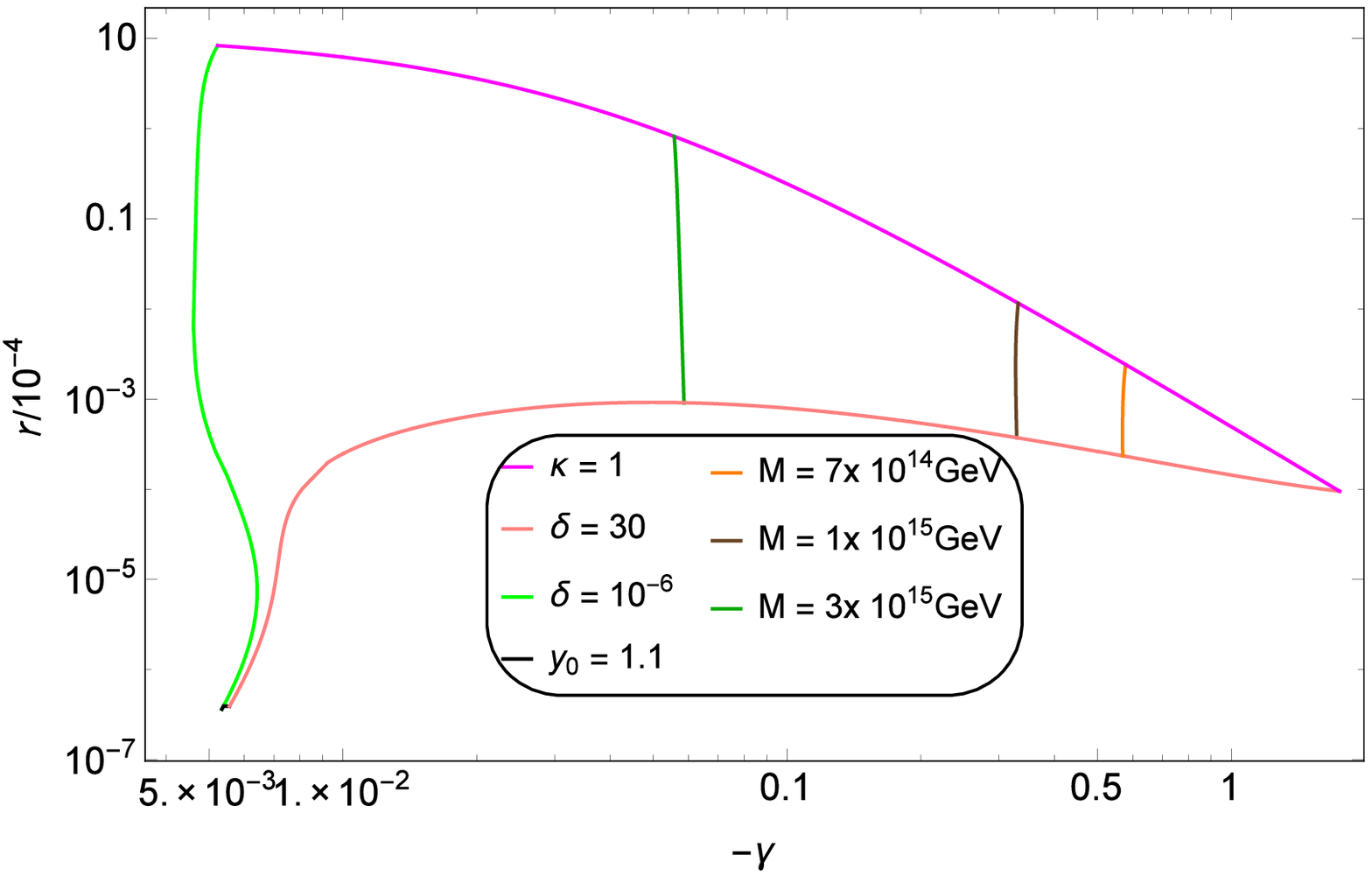}\,
\includegraphics[width=8cm]{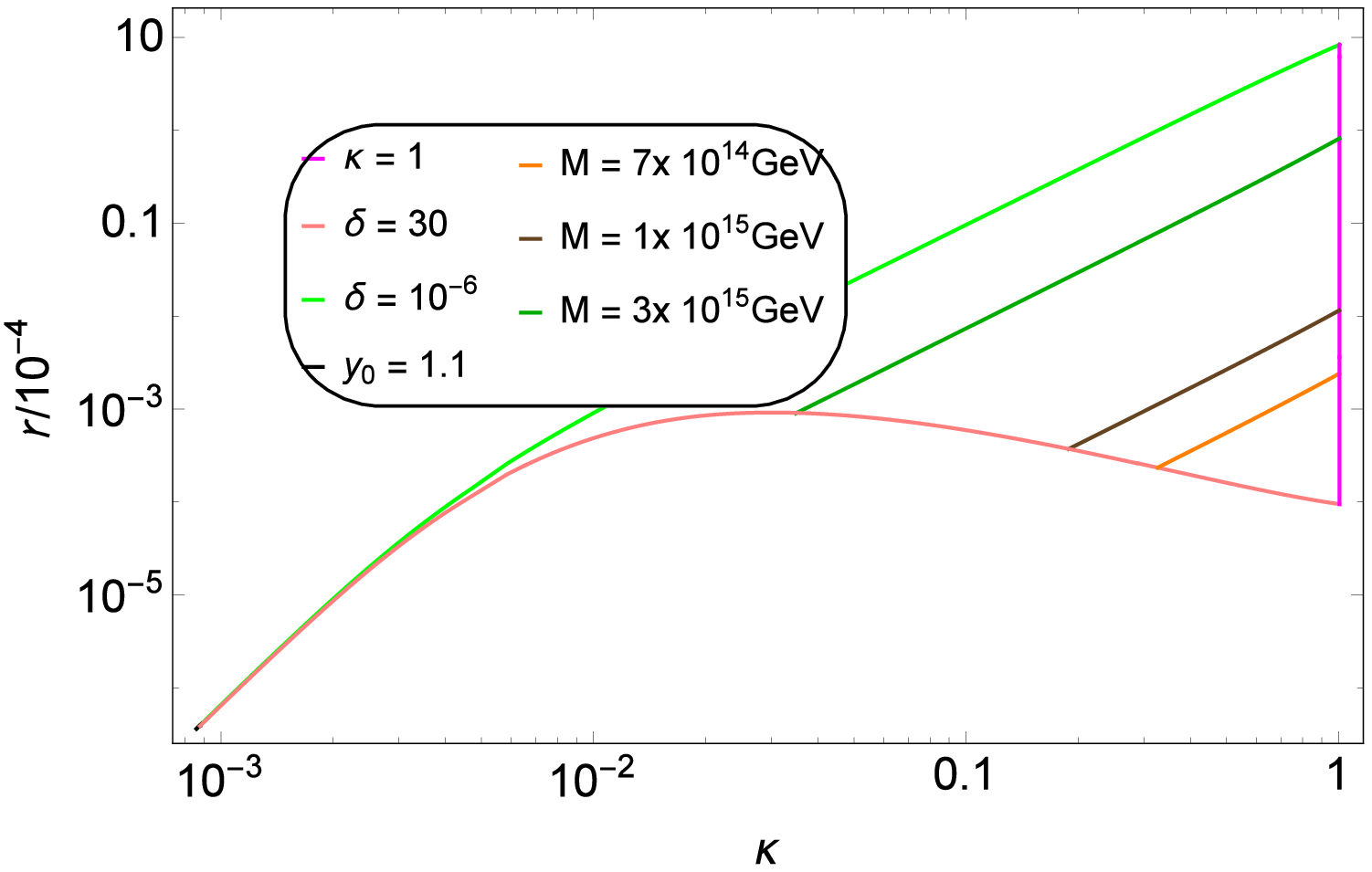}
\caption{The tensor to scalar ratio $r$ versus the dimensionless couplings $\gamma$ (left panel) and $\kappa$ (right panel). We set the scalar spectral index  $n_{s} = 0.966$ (central value of Planck's data) and the reheat temperature $T_{r}= 10^{6}$ GeV with $y_{c} = 1 $.}
\label{rgk}
\end{figure}

\begin{figure}[ht]
\includegraphics[width=8cm]{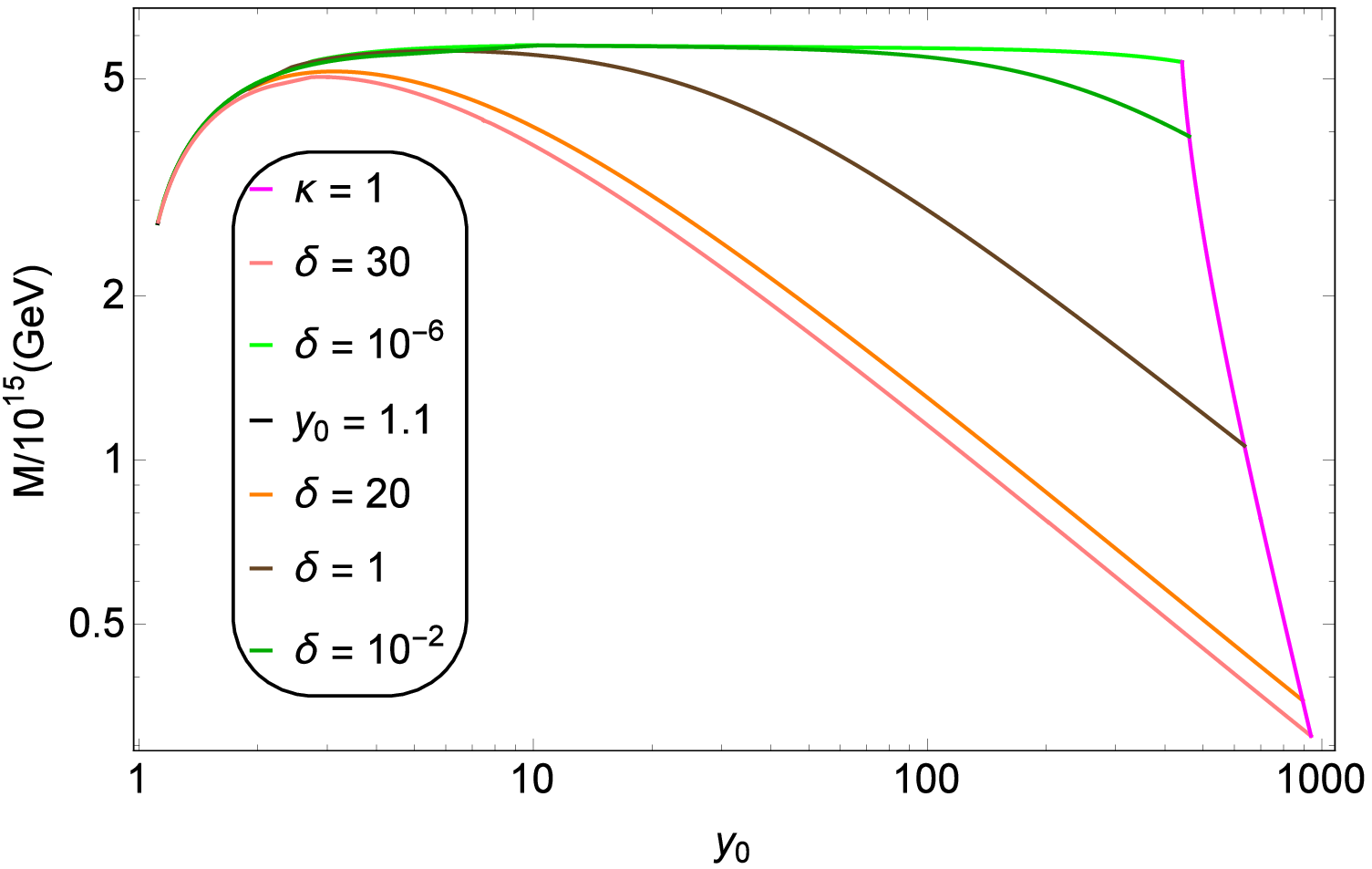}\includegraphics[width=8cm]{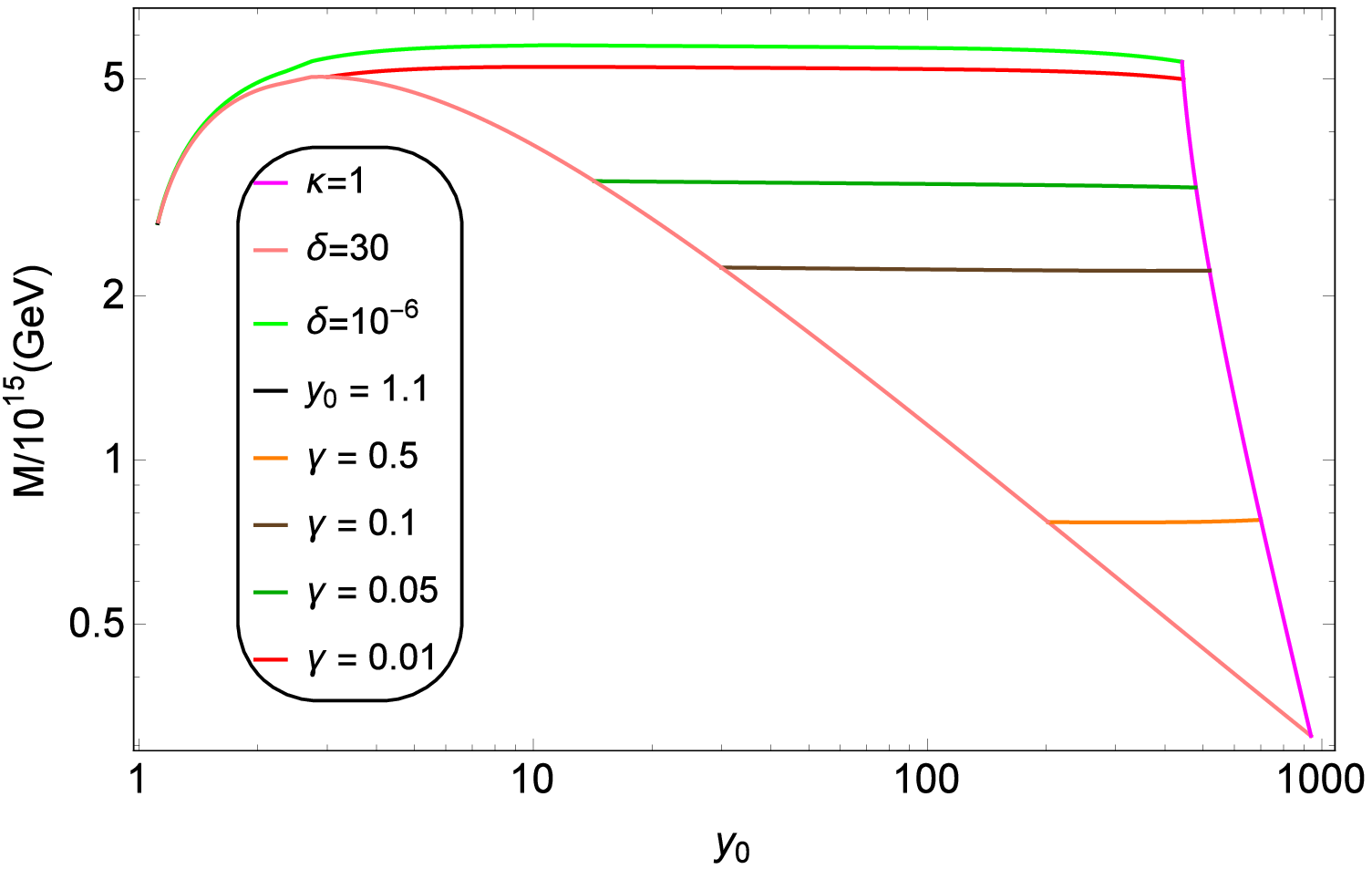}
\caption{The gauge symmetry breaking scale $M$ versus the $y_{0} = N_{0}/M$, for various values of $\delta$ (left panel) and $\gamma$ (right panel).  We set scalar spectral index  $n_{s} = 0.966$ (central value of Planck's data) and the reheat temperature $T_{r}= 10^{6}$ GeV with $y_{c} = 1 $.}
\label{My0}
\end{figure}

\begin{figure}[t]
\begin{center}
\includegraphics[width=8cm]{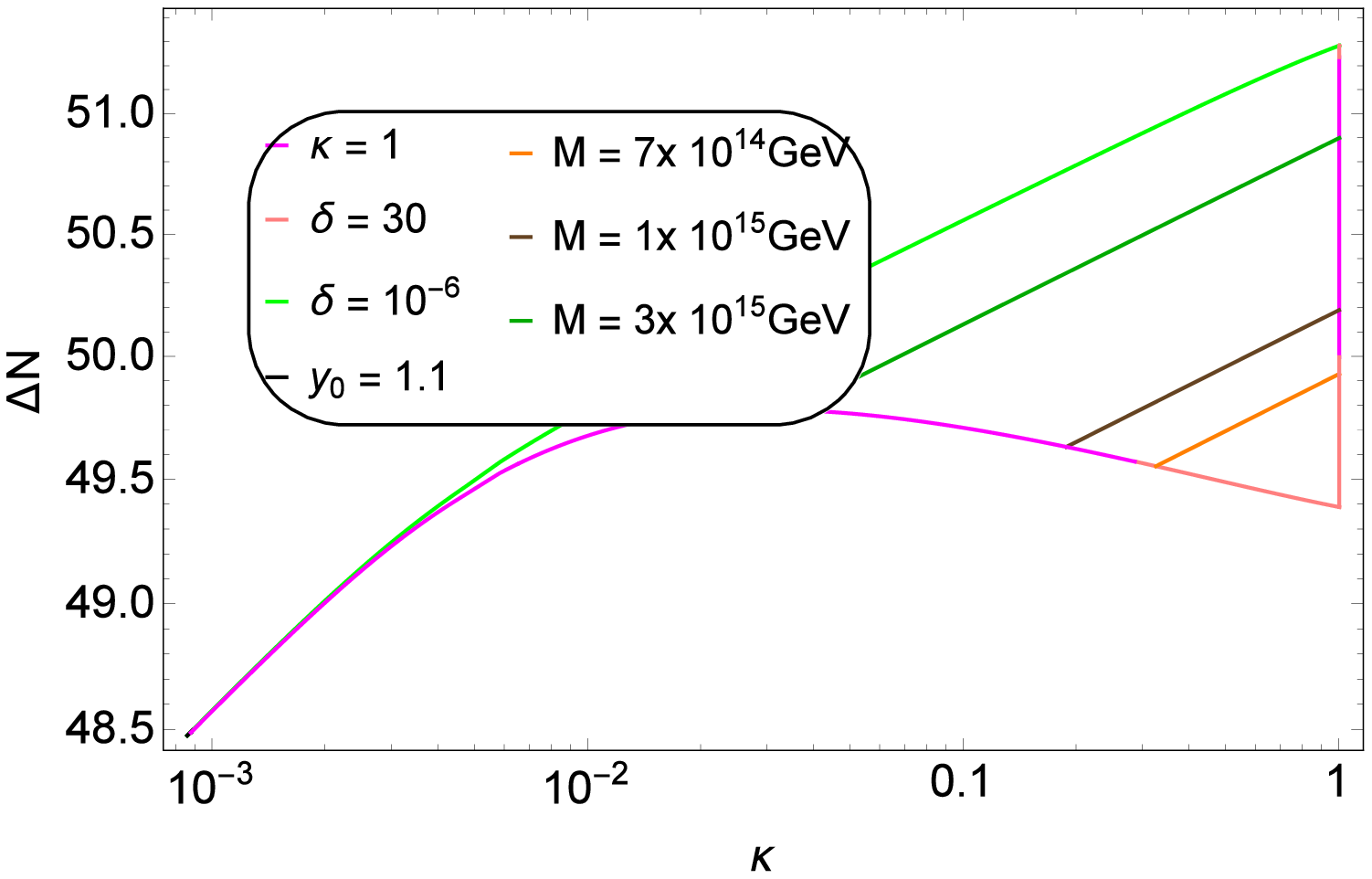}
\includegraphics[width=8cm]{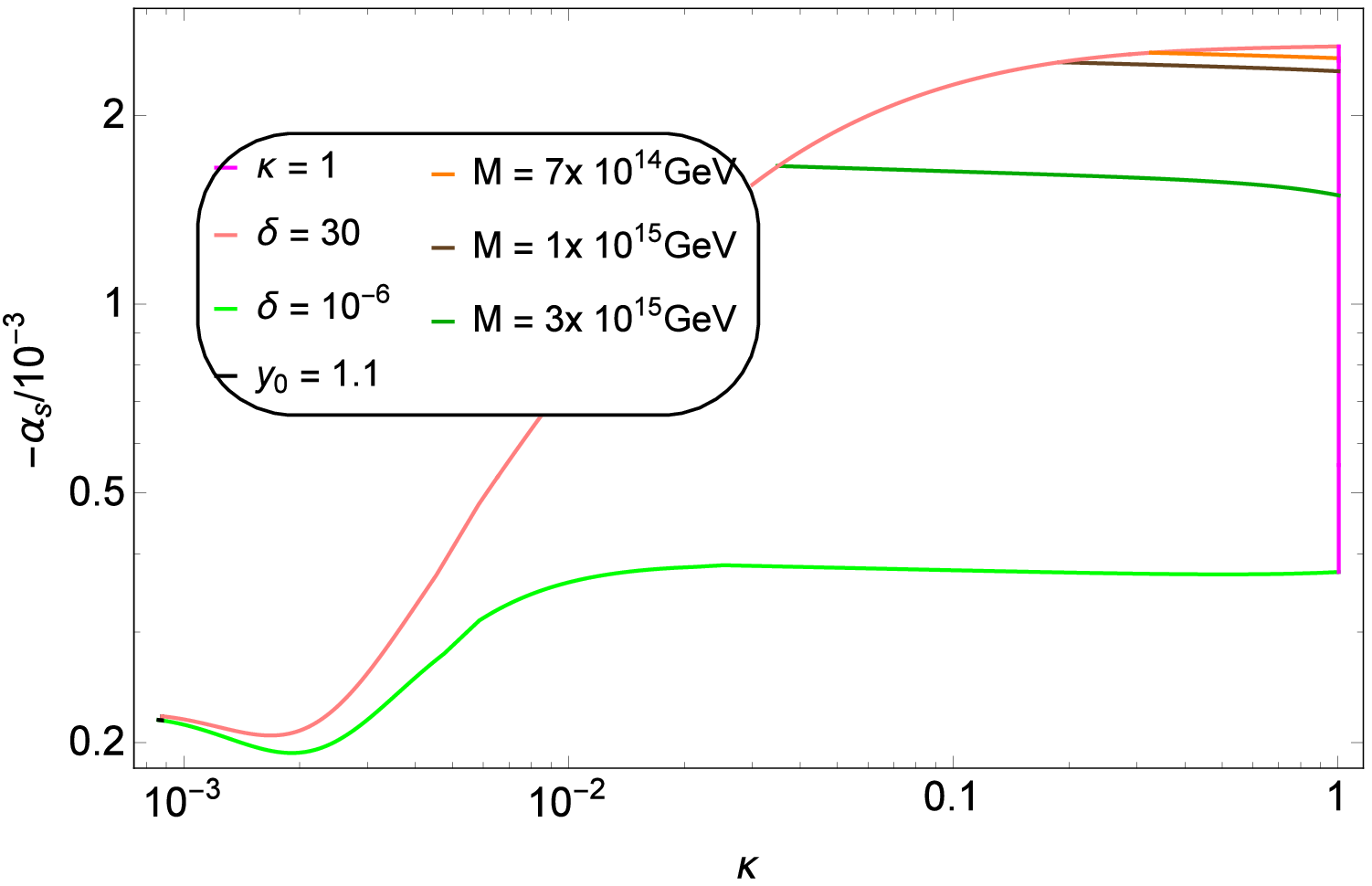}
\end{center}
\caption{The number of e-folds $\Delta N$ (left panel)  and the running of spectral index  $\alpha_s$ (right panel) versus the dimensionless coupling  $\kappa$. We set scalar spectral index  $n_{s} = 0.966$ (central value of Planck's data) and the reheat temperature $T_{r}= 10^{6}$ GeV with $y_{c} = 1 $.}
\label{dnalfa}
\end{figure}

Analogous to standard hybrid inflation the choice $\gamma < 0$, $\delta > 0$ leads to red-tilted scalar spectral index with somewhat smaller values of the tensor to scalar ratio \cite{Bastero-Gil:2006zpr,urRehman:2006hu}.  For $0 \lesssim \delta \lesssim 30$ and $-2 \lesssim \gamma \lesssim -0.005$ we obtain realistic inflationary solutions for the following range of the gauge symmetry breaking scale, $ 3 \times 10^{14}\lesssim M/ \text{GeV}\lesssim 5.4 \times 10^{15}$, the tensor to scalar ratio, $ 3 \times 10^{-11} \lesssim r \lesssim 7 \times 10^{-4} $, and the running of the scalar spectral index, $-0.00022 \lesssim dn_{s}/d \ln k \lesssim -0.0026$, corresponding to subPlanckian field values  $ 0.001\lesssim N_0/m_P \lesssim 1 $, as shown in Figs.~\ref{rgk} to \ref{dnalfa}. In the predicted range, $0.00085 \lesssim \kappa \lesssim 1$, the lower bound on $\kappa$ comes from the bound, $y_0 > y_c = 1$, whereas the upper bound $\kappa \sim 1$ is imposed on natural grounds. The upper bound on $\kappa$ also keeps the field value $N_0$ below its maximum, $N_0 \sim m_P$. A semi-analytic justification of the various limits described here is given below.


The analytic expressions for the scalar spectral index and the tensor to scalar ratio are respectively given by
\begin{eqnarray}
n_s &\simeq & 1 + 2\gamma +12\delta\left(\frac{M}{m_P}\right)^2y_0^2 + \left(\frac{\kappa^2}{8\pi^2}\right)\left(\frac{m_P}{M}\right)^2 F''(y_0), \\
r&\simeq&\left(4\gamma y_0 \left(\frac{M}{m_P}\right) + 8\delta y_0^3\left(\frac{M}{m_P}\right)^3 + \frac{\kappa^2}{4 \pi^2} F'(y_{0})\left(\frac{m_P}{M}\right)\right)^2.
\end{eqnarray}
The bounds, $\gamma \lesssim -0.005$ and $r \lesssim 0.0007 $, are achieved by taking the limit $y_0 \sim m_P/M$, $\kappa \sim 1$ and $\delta \ll 1$ in the above expressions of $n_s$ and $r$,
\begin{equation}
\gamma \lesssim \frac{1}{8\pi^2} -\frac{1-n_s}{2} \simeq -0.0043,  \quad
r  \lesssim  \left( \frac{1}{\pi^2} - 2(1-n_s)\right)^2 \simeq 0.0009,
\end{equation}
where $F''(y_0) \simeq -2/y_0^2$ and $F'(y_0) \simeq 2/y_0$ are used in the large $y_0$ limit.  Further, using Eq.~(\ref{Mkr}) we obtain $M \sim 5 \times 10^{15}$~GeV corresponding to $r \sim 0.0009$ which is also consistent with our numerical estimate. These numbers represent a good approximation of our numerical results displayed in Figs.~\ref{rgk} and \ref{My0}.

In the limit where $y_0$ approaches the waterfall point, $y_c=1$, we take $y_0 = 1.1$ to represent a limiting lowest field value with a $10 \%$ difference between $y_0$ and $y_c$. This limit also corresponds to the lower bounds on $\kappa$ and $r$ given in terms of $M$ as
\begin{equation}
\kappa \gtrsim \sqrt{\frac{2}{|F''(y_0)|}} \left( \frac{M}{m_P} \right), \quad  r \gtrsim \frac{4}{3 \pi^2 A_s|F''(y_0)|} \left( \frac{M}{m_P} \right)^6.
\end{equation}
With $y_0=1.1$, $F''(y_0) \simeq -3.46$ and $M/m_P \sim 10^{-3}$, these expressions give us the approximate lower limits,
$\kappa \gtrsim 0.0007$ and $r \gtrsim 2 \times 10^{-11}$. These analytic approximations are in good agreement with the numerical predictions  shown in  Figs.~\ref{rgk} and \ref{My0}. With Eq.~(\ref{Mkr}) and Eq.~(\ref{nne}), the predicted range of $r$ also explains the range for the number of e-folds,
$\Delta N \simeq 48.5-51.3$, as depicted in the left panel of Fig.~\ref{dnalfa}.

Finally, the running of the spectral index, $\alpha_s \equiv \frac{dn_{s}}{d\ln k}$, is given by
\begin{equation}
\alpha_s \simeq -r\frac{(1 - n_s)}{2} - \frac{3}{32}r^2 - \frac{\sqrt{r}}{8}\left(\frac{\kappa^2}{8\pi^2}F^{\prime\prime\prime}(y_0)\left(\frac{m_P}{M}\right)^3 + 24 \delta\left(\frac{M}{m_P}\right) y_0\right).
\end{equation}
In the limit $y_0 \sim 0.1 m_P/M$, $F^{\prime\prime\prime}(y_0) \simeq 4/y_0^3$, $\kappa \sim 1$, $r\sim 10^{-8}$ and $\delta \sim 30$ we obtain 
\begin{equation}
\alpha_s \simeq - 3\sqrt{r}\,\delta\left(\frac{M}{m_P}\right) y_0 \simeq -0.002,
\end{equation}
while in the limit $y_0 =1.1$, $F'''(y_0) \simeq 27$, $\kappa \sim 0.0009$, $r \sim 3 \times 10^{-11}$ and $M/m_P \sim 10^{-3}$ we obtain
\begin{equation}
\alpha_s \simeq - \frac{\sqrt{r}}{8}\left(\frac{\kappa^2}{8\pi^2}F^{\prime\prime\prime}(y_0)\left(\frac{m_P}{M}\right)^3\right) \simeq -0.0002.
\end{equation}
These analytic approximations are in good agreement with the numerical estimates shown in the right panel of Fig.~\ref{dnalfa}.
\subsection{The Case with $\gamma > 0$, $\delta < 0$}
\begin{figure}[t]
\begin{center}
\includegraphics[width=8cm]{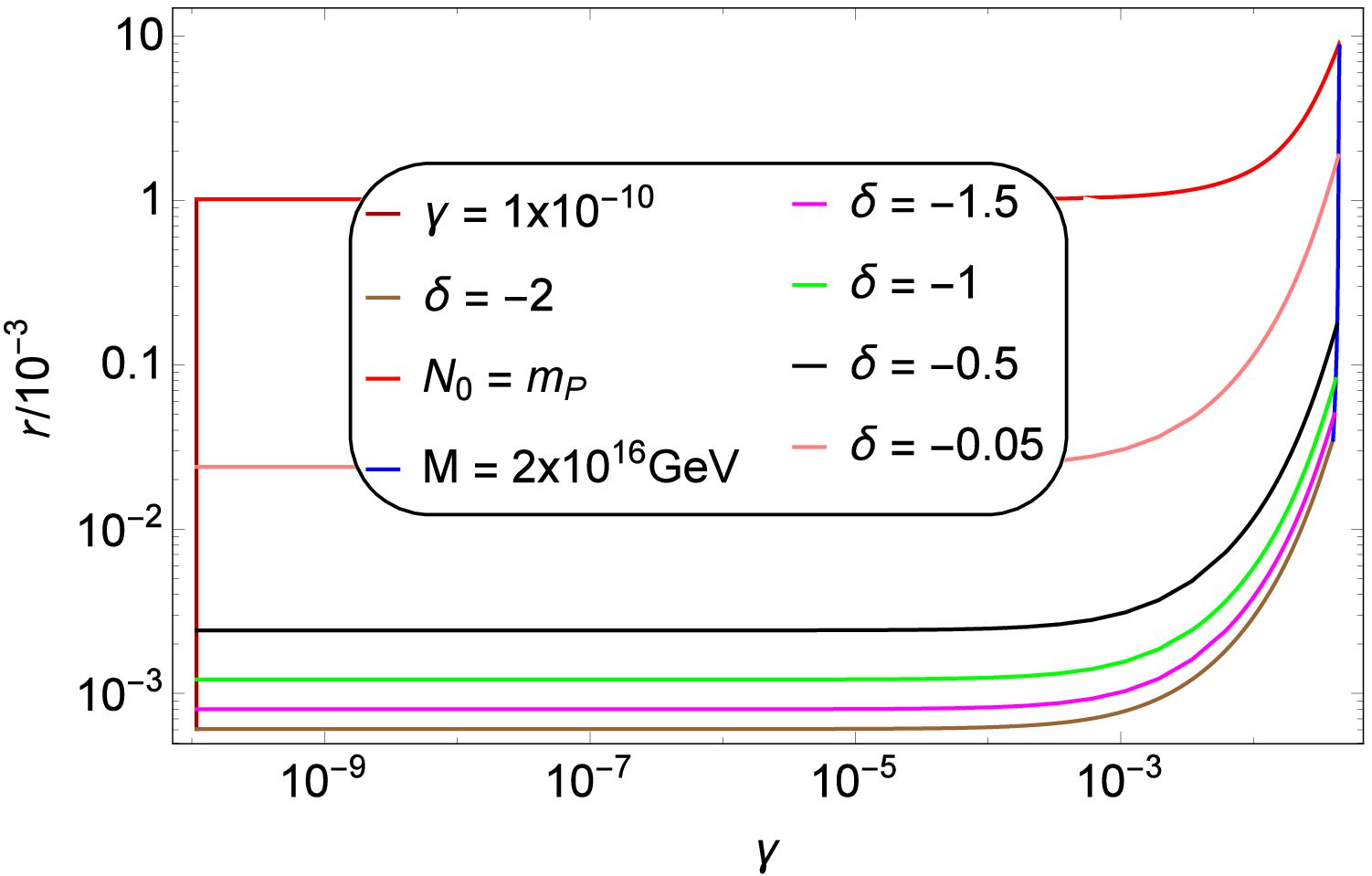}
\includegraphics[width=8cm]{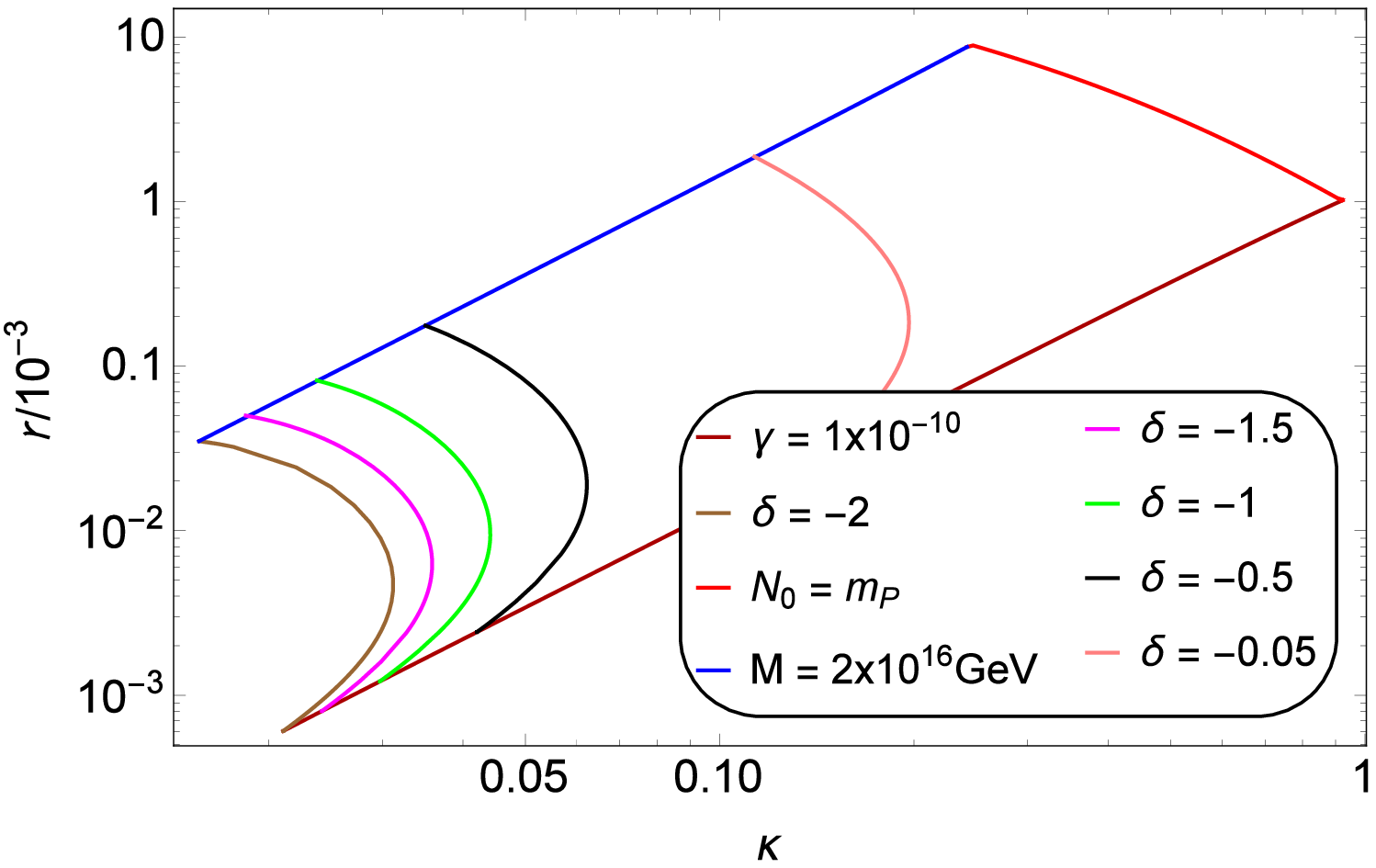}
\end{center}
\caption{The tensor to scalar ratio $r$ versus the dimensionless couplings $\gamma$ (left panel) and $\kappa$ (right panel). We set the scalar spectral index  $n_{s} = 0.966$ (central value of Planck's data) and the reheat temperature $T_{r}= 10^{6}$ GeV with $y_{c} = 1 $.}  \label{rgk2}
\end{figure}

\begin{figure}[t]
\begin{center}
\includegraphics[width=8cm]{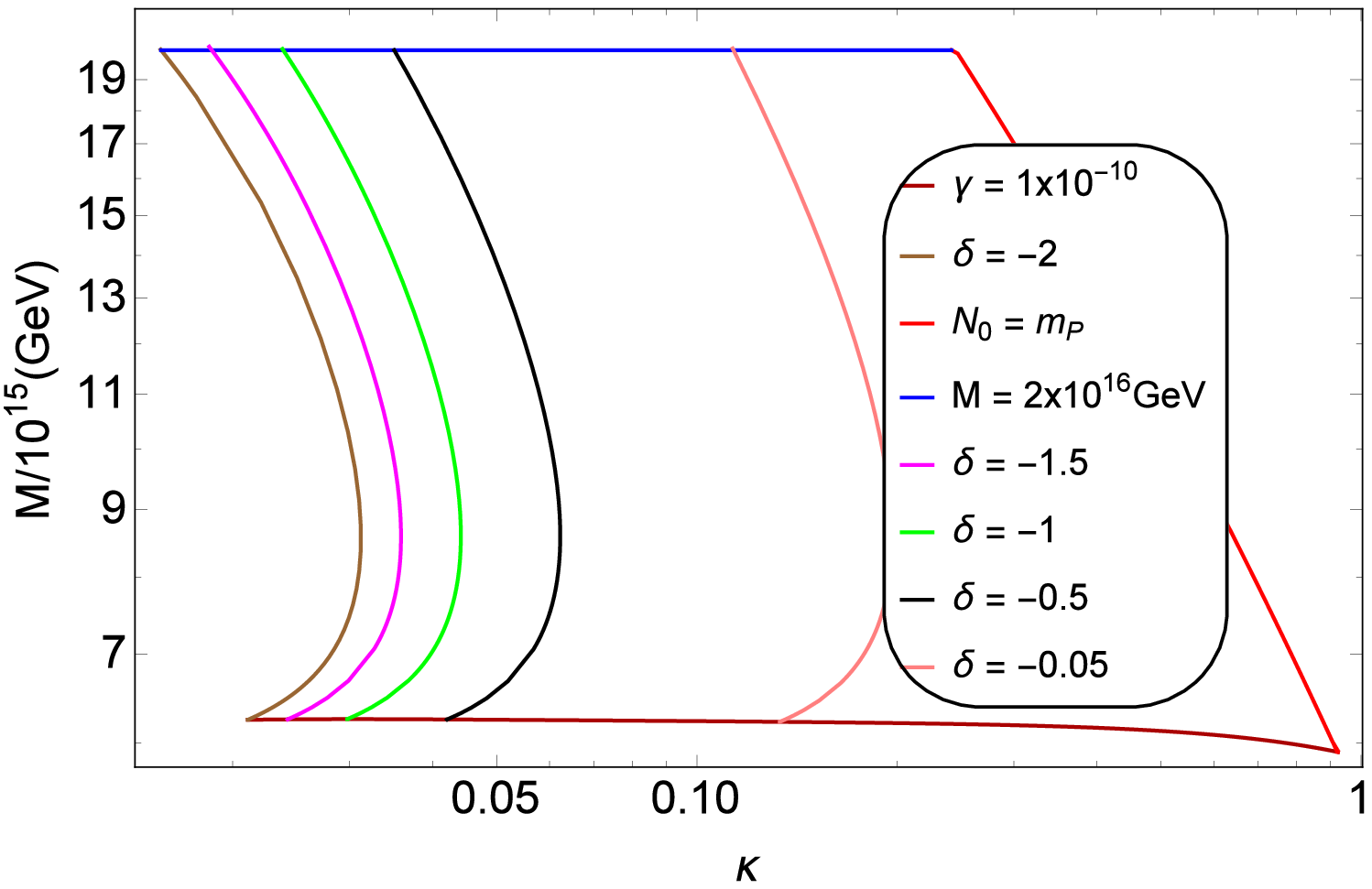}
\includegraphics[width=8cm]{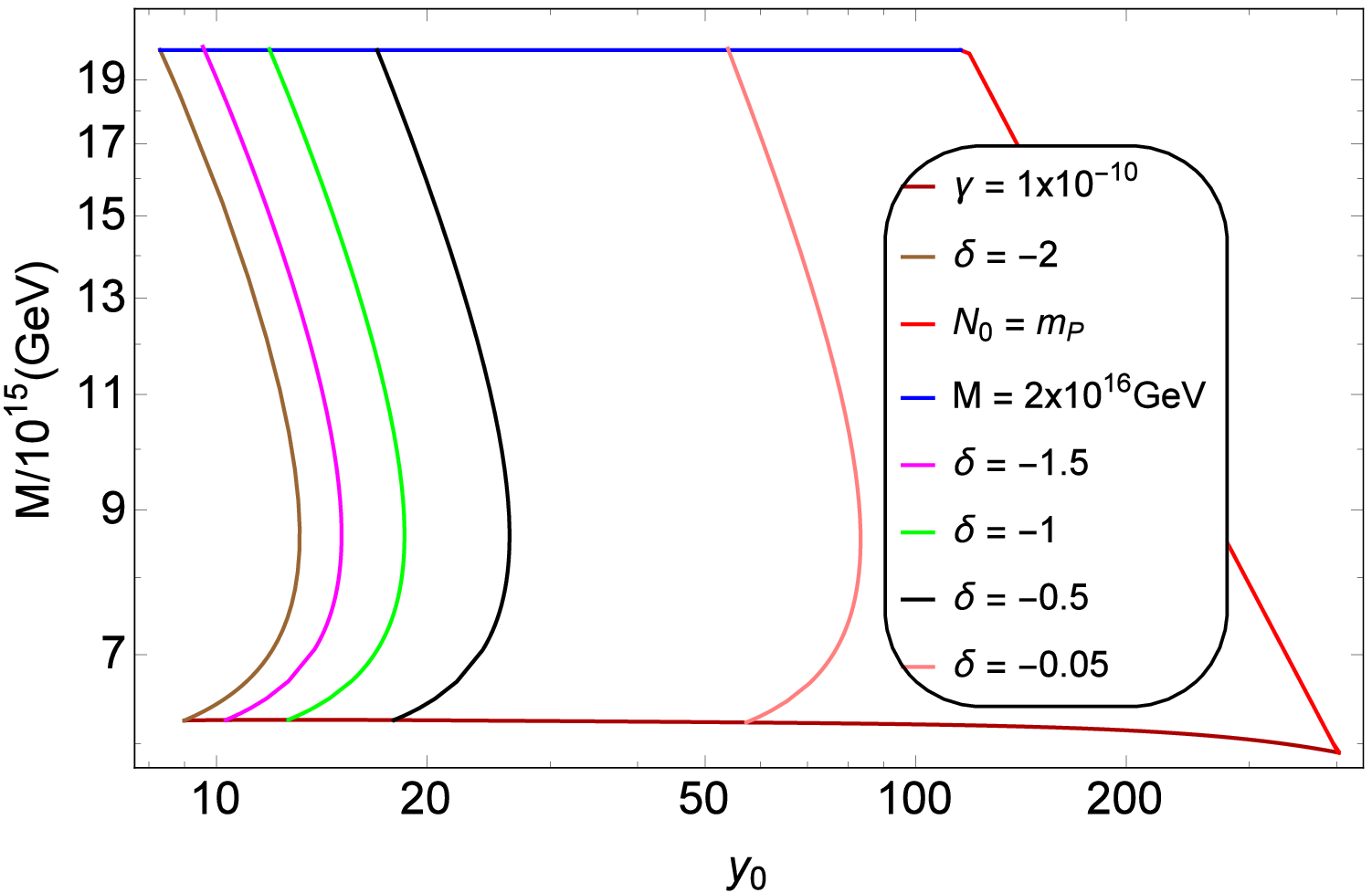}
\end{center}
\caption{The gauge symmetry breaking scale $M$ versus $\kappa$ (left panel) and $y_{0} = N_{0}/M$ (right panel), for various values of $\delta$ (left panel) and $\gamma$ (right panel).  We set scalar spectral index  $n_{s} = 0.966$ (central value of Planck's data) and the reheat temperature $T_{r}= 10^{6}$ GeV with $y_{c} = 1 $.}  \label{Mgk2}
\end{figure}

\begin{figure}[t]
\begin{center}
\includegraphics[width=8cm]{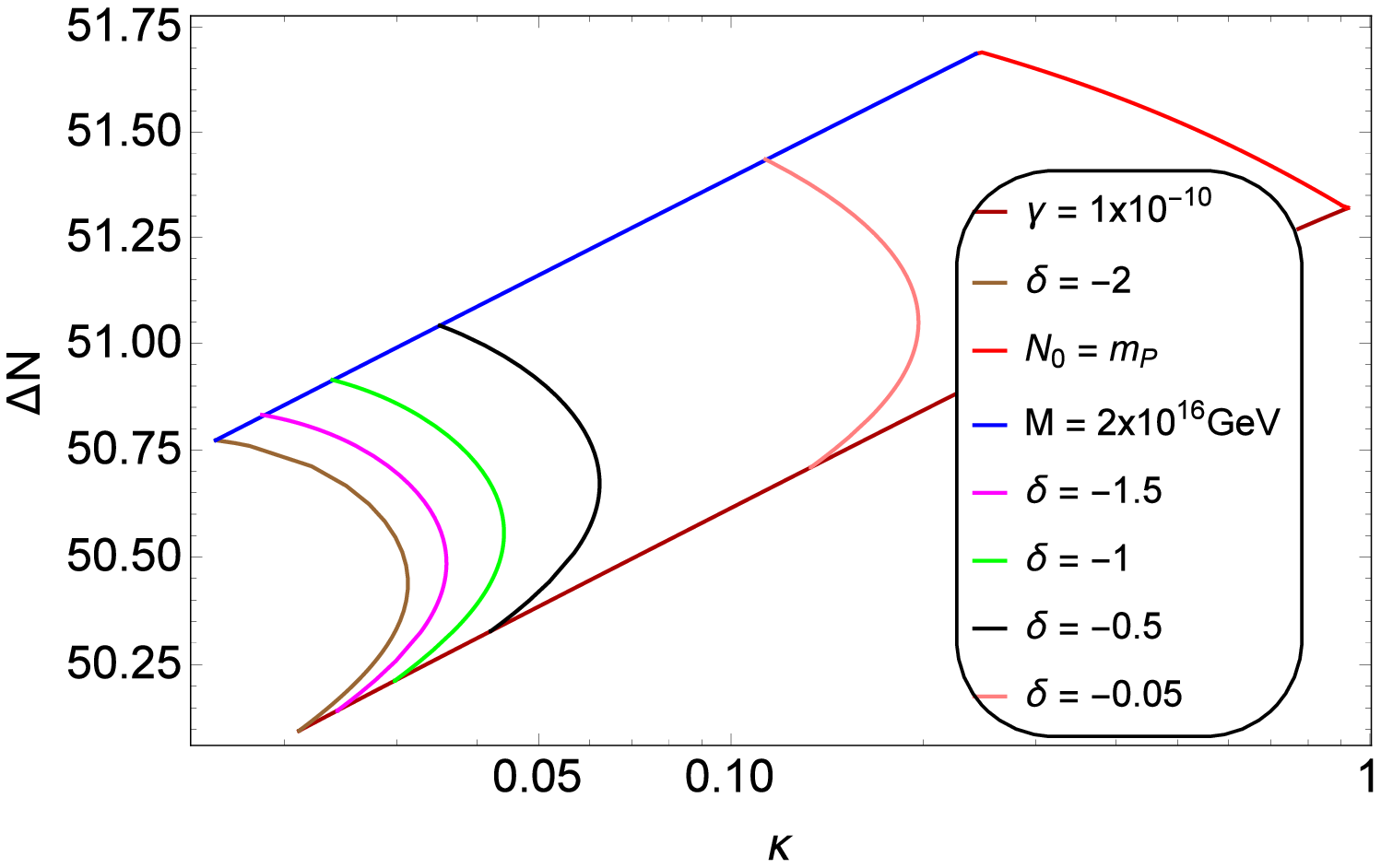}
\includegraphics[width=7.6cm]{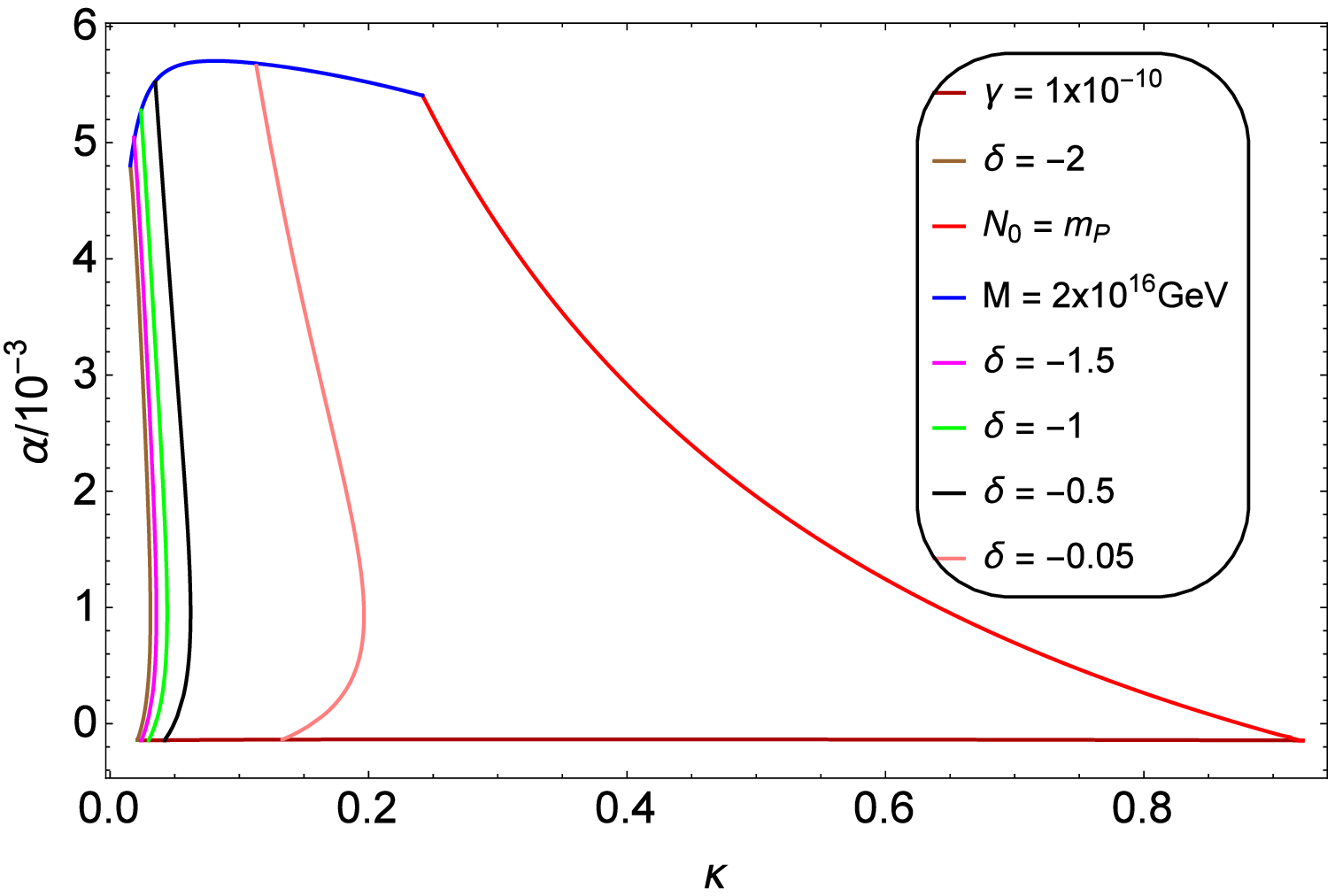}
\end{center}
\caption{The number of e-folds $\Delta N$ (left panel)  and the running of spectral index  $\alpha_s$ (right panel) versus the dimensionless coupling  $\kappa$. We set scalar spectral index  $n_{s} = 0.966$ (central value of Planck's data) and the reheat temperature $T_{r}= 10^{6}$ GeV with $y_{c} = 1 $.}  \label{Nak2}
\end{figure}

The choice $\gamma > 0$, $\delta < 0$, can lead to large $r$ solutions, $r \gtrsim 10^{-3}$, which lie within the detectable range of primordial gravitational waves of future experiments
\cite{PRISM:2013ybg,Matsumura:2013aja,Kogut:2011xw,CORE:2016ymi}. For SUSY hybrid inflation, these solutions with observable primordial gravitational waves have been explored in \cite{Rehman:2010wm} in detail. For model parameters lying in their range $-2 \lesssim \delta \lesssim -0.0012$, $0 \lesssim \gamma \lesssim 0.044$ and $0.0155 \lesssim \kappa \lesssim 0.92 $, we obtain realistic inflationary solutions  $ 5.9 \times 10^{15}\lesssim M/ \text{GeV}\lesssim 2 \times 10^{16}$, $ 6 \times 10^{-7} \lesssim r \lesssim 0.009 $, and $-0.00014 \lesssim \alpha_s \lesssim 0.005$, corresponding to subPlanckian field values  $ 0.023 \lesssim N_0/m_P \lesssim 1 $, as shown in Figs.~\ref{rgk2} to \ref{Nak2}.

For a semi-analytic justification of the various limits shown in the Figs.~\ref{rgk2} to \ref{Nak2} we consider the following expressions for $n_s$ and $r$:
\begin{eqnarray}
n_s &\simeq & 1 + 2\gamma +12\delta\left(\frac{N_0}{m_P}\right)^2 - \left(\frac{\kappa^2}{4\pi^2}\right)\left(\frac{m_P}{N_0}\right)^2,  \label{ns22}  \\
r  &\simeq&  \left(4\gamma \left(\frac{N_0}{m_P}\right) + 8 \delta \left(\frac{N_0}{m_P}\right)^3 + \frac{\kappa^2}{2 \pi^2} \left(\frac{m_P}{N_0}\right)\right)^2,  \label{r22}
\end{eqnarray}
where $F''(y_0) \simeq -2/y_0^2$ and $F'(y_0) \simeq 2/y_0$ are assumed in the predicted range of $y_0 \gtrsim 10$. The upper bound on the tensor to scalar ratio, $r \lesssim 0.009 $, is achieved by taking the limit $N_0 \sim m_P$ in the above expressions of $n_s$ and $r$,
\begin{equation}
r  \lesssim  \left( 4\gamma + 8 \delta \right)^2,    \quad
\delta \sim - \frac{\gamma}{6} - \frac{1-n_s}{12}.
\end{equation}
With $\gamma \sim 0.044$ we obtain $\delta \sim 0.01$ and $r \lesssim 0.01$. Assuming $M \sim 2 \times 10^{16}$~GeV, this bound on $r$ corresponds to $\kappa \sim 0.25$ via Eq.~(\ref{Mkr}) which, in turn, justifies the neglect of radiative corrections in deriving the above equation. The lower bound on $r$ is achieved in the limit $\gamma \sim 0$ with $\delta = - 2$,
\begin{equation}
r  \lesssim  \left( 8 \, \delta  \left( \frac{N_0}{m_P}\right)^4 - \frac{\kappa^2}{4\pi^2} \right)^2\left( \frac{m_P}{N_0}\right)^2,    \quad
\left( \frac{N_0}{m_P}\right)^2 \sim  \frac{1-n_s + \sqrt{(1-n_s)^2 + \frac{24\,\kappa^2}{\pi^2}}}{48}.
\end{equation} 
With $\kappa \sim 0.021$ we obtain $N_0/m_P \sim 0.03$, $r \gtrsim 6 \times 10^{-7}$ and  $M \sim 6.2 \times 10^{15}$~GeV via Eq.~(\ref{Mkr}). All pf these approximations are consistent with our numerical results displayed in Figs.~\ref{rgk2} and \ref{Mgk2}. Using Eq.~(\ref{Mkr}) and Eq.~(\ref{nne}), the predicted range of $r$ leads to the range for the number of e-folds,
$50.1 \lesssim \Delta N \lesssim 51.7 $, as depicted in the left panel of Fig.~\ref{Nak2}.

The intersection of the curves with $N_0 \sim m_P$ and $\gamma \sim 0$ gives rise to the following relations,
\begin{equation}
\delta \sim  -\frac{1-n_s}{12} + \frac{\kappa^2}{48 \, \pi^2},  \quad  r \sim \left( 8 \,\delta + \frac{\kappa^2}{2 \pi^2} \right)^2.
\end{equation} 
With $\kappa \sim 0.92$ we obtain $\delta \sim -0.001$ and $r \sim 0.001$ consistent with our numerical estimates. Similarly, the intersection of the curves with $M \sim 2 \times 10^{16}$~GeV and $\delta = -2$ gives rise to $\kappa \sim 0.0155$, $\gamma \sim 0.04$ and $N_0/m_P \sim 0.0001$ for $r \sim 0.000035$ by employing Eq.~(\ref{Mkr}), Eq.~(\ref{ns22}) and Eq.~(\ref{r22}).

Finally, the running of the spectral index, $\alpha_s$, is given by
\begin{equation}
\alpha_s \simeq -r\frac{(1 - n_s)}{2}  - \frac{3}{32}r^2 - \frac{\sqrt{r}}{8}\left(\frac{\kappa^2}{2\pi^2}\left(\frac{m_P}{N_0}\right)^3 + 24 \, \delta\left(\frac{N_0}{m_P}\right)\right).
\end{equation}
In the large $r$ limit with $N_0 \sim m_P$  we finds 
\begin{equation}
\alpha_s \simeq - 3\sqrt{r}\,\delta \simeq 0.003,
\end{equation}
and in the small $r$ limit with $\delta = -2$ and $\gamma \sim 0$, 
\begin{equation}
\alpha_s \simeq - \frac{\sqrt{r}}{8}\left(\frac{\kappa^2}{2\pi^2}\left(\frac{m_P}{N_0}\right)^3 - 48 \left(\frac{N_0}{m_P}\right)\right) \simeq -0.0001.
\end{equation}
These analytic approximations are in good agreement with the numerical estimates shown in Fig.~\ref{Nak2}.
\section{Reheating and Non-Thermal Leptogenesis}
The reheating in the current model proceeds in analogy with the $Z_4$ sneutrino model considered in \cite{Antusch:2004hd} and the observed baryon asymmetry is generated via non-thermal leptogenesis \cite{Lazarides:1991wu}. We take the inflaton $N$ to be the lightest sneutrino field $N_{1}$, decaying through the Yukawa coupling, $y_{ij}^{\nu} L_{i} H_{u} N_{j}^c $
into slepton and Higgs or into lepton and Higgsino with a decay width given by  
\begin{align}\label{eq:22}
&\Gamma_{N_{1}} \simeq \frac{y_{\nu}^2}{4\pi} M_{R}^I = \frac{y_{\nu}^2}{4\pi}\left(\frac{\kappa M}{y_c^2}\right).
\end{align}
Here, $ y_{\nu}^2 \equiv  (y_{\nu}y_{\nu}^{\dagger})_{11}$ and using Eq.~(\ref{majmass}), the  mass of the sneutrino inflaton $N_{1}$ can be expressed in terms of $y_c$ as
\begin{align}\label{eq:21}
&M_R^{I} = \left(\lambda_{11}+\tilde \lambda\right)\frac{ M^{2}}{M_{*}} = \left(2 \lambda \right)\frac{ M^{2}}{M_{*}} =\left(\frac{\kappa M}{ y_c^2}\right).
\end{align}
With $y_c = 1$ this implies that $M_R^I \lesssim M$ for $\kappa \lesssim 1$.

\begin{figure}[tp]
\includegraphics[width=8cm]{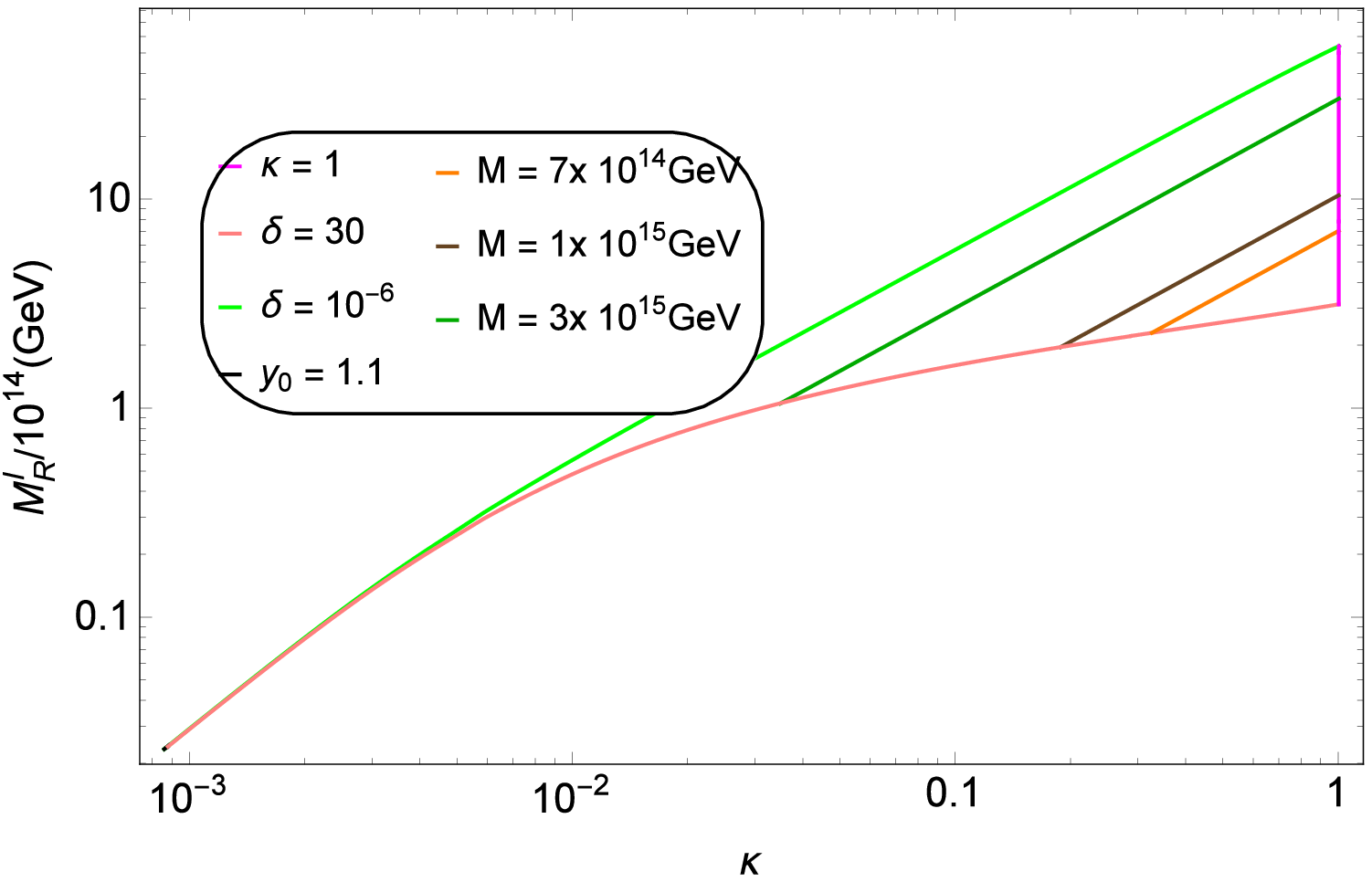}
\includegraphics[width=8cm]{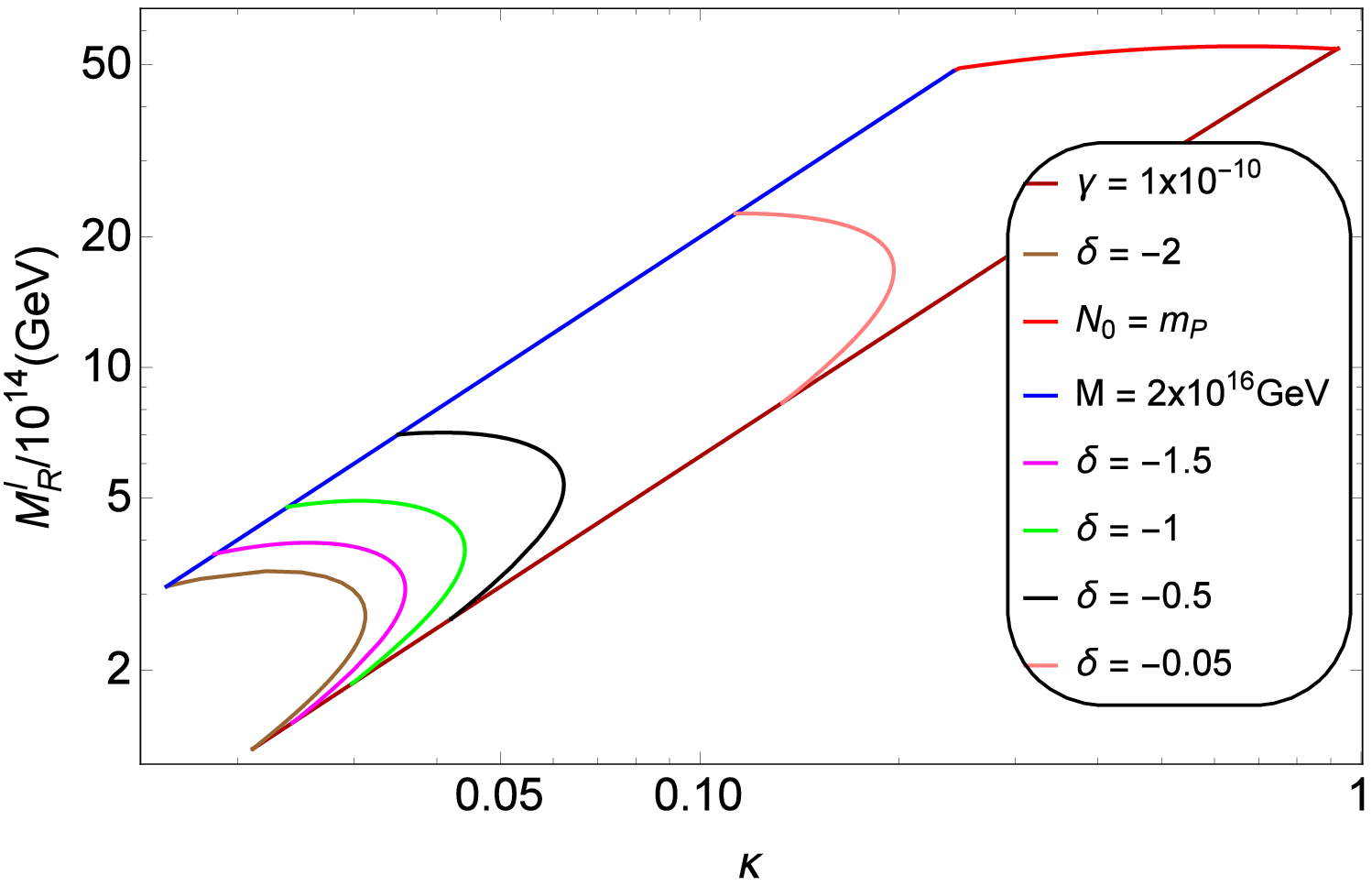}
\includegraphics[width=8cm]{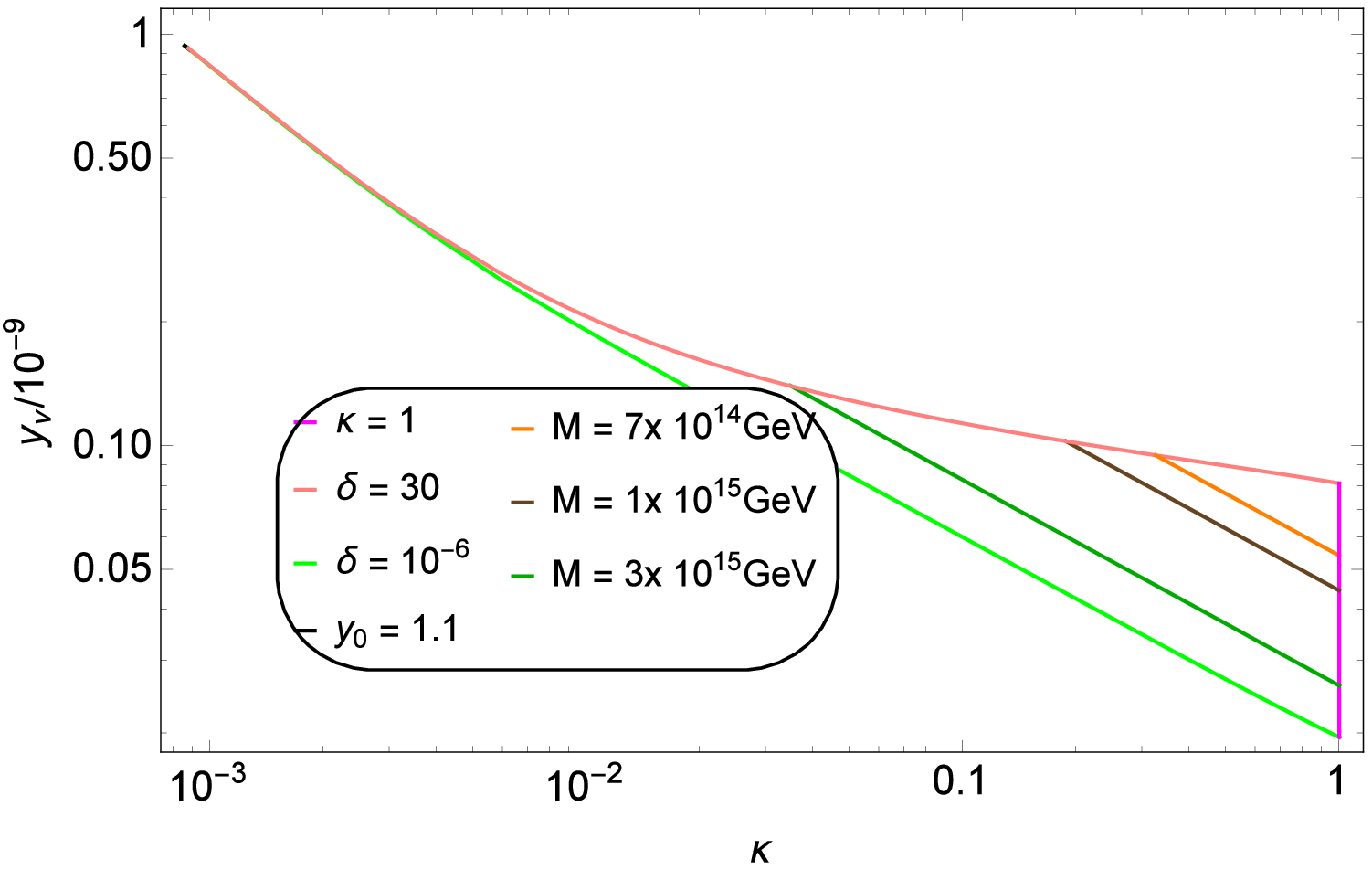}
\includegraphics[width=8cm]{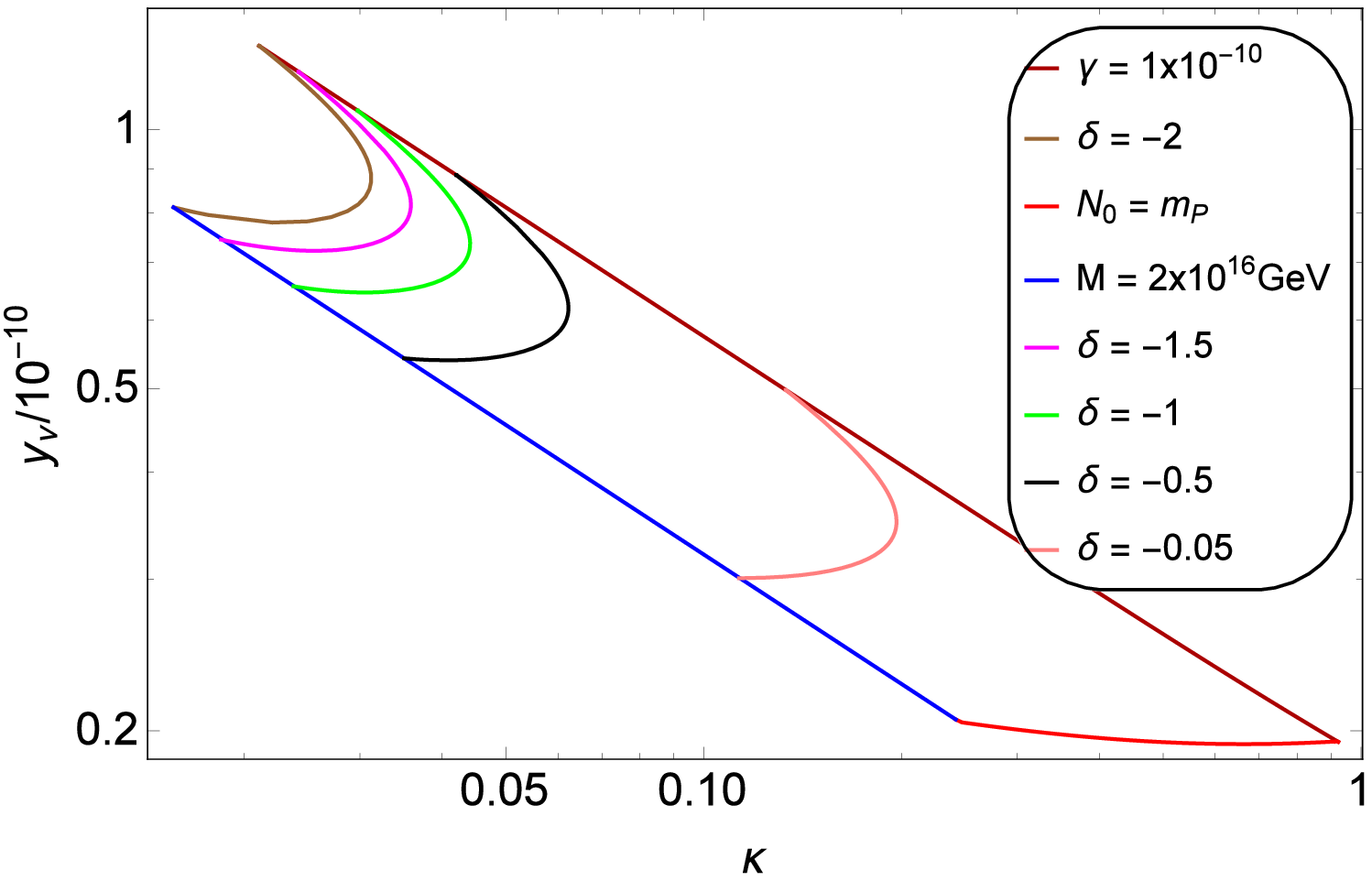}
\caption{ The mass of the (s)neutrino inflaton $M_{R}^I$ (left panels) and  the neutrino Yukawa coupling $y_{\nu}$ (right panels) versus the coupling $\kappa$. We set the scalar spectral     index  $n_{s} = 0.966$ (central value of Planck's data) and temperature $ T_{r} = 10^{6}$ GeV with $y_c = 1$.}
\label{MrTr}
\end{figure}

In general, both the inflaton and waterfall field can be relevant for reheating and leptogenesis. However, for the inflaton to play a dominant role in reheating and leptogenesis it has to decay later than the $\Phi$ field which decays earlier predominantly into the heavier neutrino $N_{2}$ (or $N_{3}$).  To satisfy this requirement the Higgs decay rate, $\Gamma_{\Phi}$, has to be larger than the inflaton decay rate, $\Gamma_{N_1}$. This leads to the following bound on $y_{\nu}$
\begin{eqnarray}
y_{\nu}^2 \equiv  (y_{\nu}y_{\nu}^{\dagger})_{11} \ll (y_{\nu}y_{\nu}^{\dagger})_{22,33} \left(\frac{M_R^{(2,3)}}{M_R^I}\right),
\end{eqnarray}
where $M_R^{(2,3)}$ are the masses of the heavier neutrinos $N_{(2,3)}$. This bound is automatically satisfied in our model for the numerical data displayed in Fig.~\ref{MrTr}.
After inflation, the universe reheats via inflaton decay to a reheat temperature given by
\begin{eqnarray}
T_{r} \simeq \left( \frac{90}{g_* \pi^{2}} \right)^\frac{1}{4}\sqrt{\Gamma_{N_{1}}m_{P}},
\end{eqnarray}
with $g_* =228.75$. 

The lepton asymmetry generated by the inflaton decay is partially converted into baryon asymmetry through the sphaleron process \cite{Kuzmin:1985mm,Fukugita:1986hr,Khlebnikov:1988sr}. The washout factor of lepton asymmetry can be suppressed by assuming  $M_{R}^I\gg T_{r}$. The observed baryon asymmetry is evaluated in term of the lepton asymmetry factor, $\varepsilon_L$,
\begin{eqnarray}
\frac{n_{B}}{n_{\gamma}}\simeq -1.84 \,\varepsilon_L  \frac{T_{r}}{M_{R}^I}\delta_{eff},
\end{eqnarray} 
where $\varepsilon_L$ is bounded by 
\begin{eqnarray}
(-\varepsilon_L)  \lesssim \frac{3}{8\pi}  \frac{\sqrt{\Delta m_{31}^{2}} M_{R}^I}{\langle H_{u}\rangle^{2}},
\end{eqnarray} 
assuming hierarchical neutrino masses. Here, the atmospheric neutrino mass squared difference is $\Delta m_{31}^{2}\approx 2.6 \times 10^{-3}$ eV$^{2} $  and $\langle H_{u}\rangle \simeq 174$ GeV in the large $\tan\beta$ limit. Finally, the bound on $\varepsilon_L$ translates into a bound on the reheat temperature $T_r \gtrsim 10^6$ GeV for the observed baryon-to-photon ratio  $n_\mathrm{B} /n_\gamma =  (6.12\pm 0.04) \times 10^{-10}$ \cite{ParticleDataGroup:2020ssz}. Thus, the reheat temperature is allowed to be small enough to avoid the gravitino problem. We set $T_r = 10^6$ GeV in all numerical work and obtain $ 2.3 \times 10^{12} \lesssim M_R^I / \text{GeV} \lesssim 5.3\times 10^{15}$ ($ 1.3 \times 10^{14} \lesssim M_R^I / \text{GeV} \lesssim 5.5 \times 10^{15}$) and $ 2\times 10^{-11} \lesssim y_{\nu} \lesssim 9.4 \times10^{-10}$ ($ 1.9 \times 10^{-11} \lesssim y_{\nu} \lesssim 1.3 \times10^{-10}$) for $\gamma < 0$, $\delta > 0$  ($\gamma > 0$, $\delta < 0$), as shown in the Fig.~\ref{MrTr}.
\section{Metastable Cosmic Strings and Stochastic Gravitational Wave Background}
Cosmic strings arise from the breaking of  $U(1)_{B-L}$ at the end of inflation. The various experimental bounds on these strings are usually described in terms of the string tension, $G\mu_s$, where $G = 1/8 \pi m_P^2$ is Newton's gravitational constant and $\mu_s$ denotes the mass per unit length of the string. 
For our case, $\mu_s$ can be written in term of $M$ as \cite{Hill:1987ye}
\begin{equation}\label{cosmicmu}
\mu_s= 2\pi M^2 \epsilon(\beta), \quad \epsilon(\beta) \approx \left\{ \begin{array}{ll}
1.04\beta^{0.195}, & \mbox{ $ \beta > 10^{-2},$} \\
\frac{2.4}{\log[2/\beta]}, & \mbox{ $ \beta < 10^{-2}$},\end{array} \right.  
\end{equation}
where $\beta = \frac{\kappa^2}{2g^2}$ with $g=0.7$ for MSSM. The CMB bound on the cosmic string tension reported by Planck 2018 \cite{Ade:2013xla,Ade:2015xua} is
\begin{eqnarray}
G \mu_s \lesssim 2.4 \times 10^{-7}.
\end{eqnarray}
For a stable cosmic string network this bound is a bit restrictive for the inflationary parameter space discussed above, but it can be relaxed for a metastable cosmic string network. 
More importantly, one may argue that metastable strings are naturally expected in the context of grand unification. Below, we discuss such a possibility for metastable cosmic string network which may also provide a possible explanation for the recent evidence of stochastic gravitational-wave background reported by the NANOGrav 12.5-year data set \cite{Arzoumanian:2020vkk}.

\begin{figure}[tp]
\includegraphics[width=7.8cm]{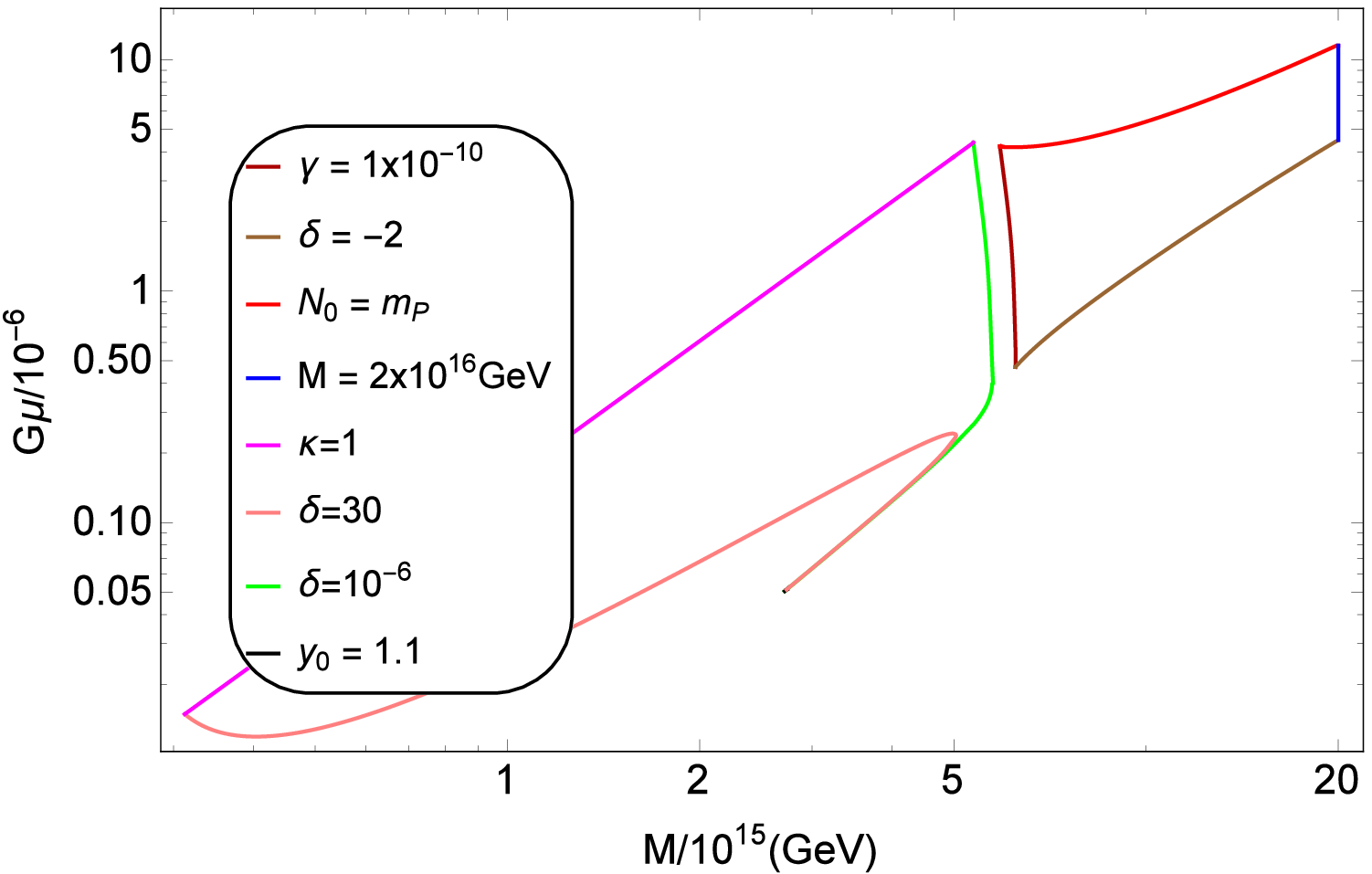}
\includegraphics[width=8.1cm]{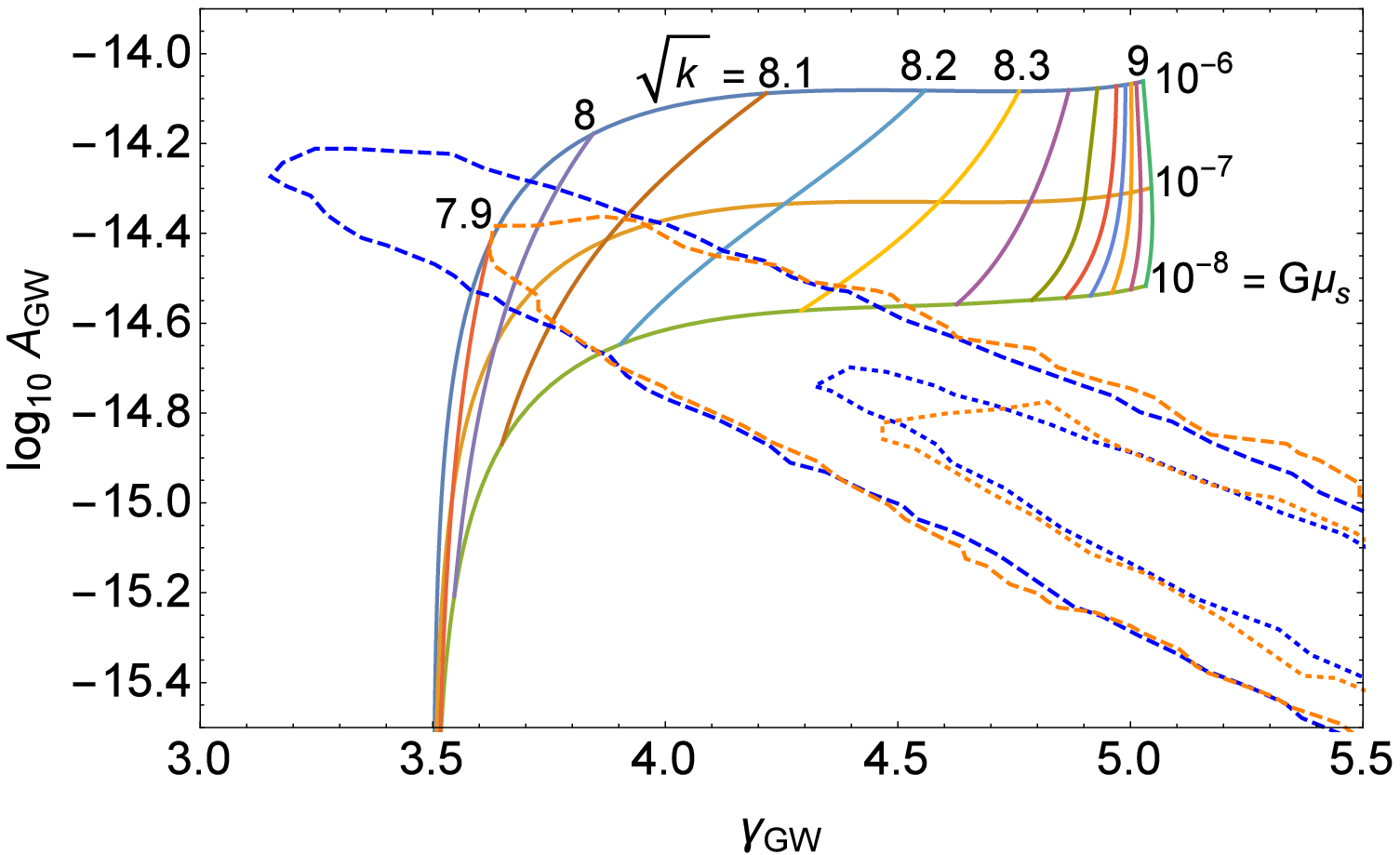}
\caption{The predicted range of $G\mu_s$ versus $M$ is shown in the left panel. The amplitude, $A_{GW}$, versus spectral index, $\gamma_{GW}$, for various values of cosmic string tension, $G\mu_s=10^{-6},\,10^{-7}, \,10^{-8}$, is shown in the right panel together with the $1$-$\sigma$ (dotted) and $2$-$\sigma$(dashed) 5-frequency power law (orange) and broken power law (blue) likelihood contours from the NANOGrav 12.5-yr 5-frequency chain data \cite{Arzoumanian:2020vkk}.}
\label{nano}
\end{figure}

The NANOGrav collaboration has presented their data \cite{Arzoumanian:2020vkk} using the following form of the energy density in gravitational waves 
\begin{equation}
\Omega_{GW} (f) = \frac{2 \pi^2 f_{\text{yr}}^2}{3 H_0^2 A_{GW}^2}  \left(\frac{f}{f_{\text{yr}}}\right)^{5 - \gamma_{GW}}\,,
\end{equation}
where $f_{\text{yr}} = 1/\text{year}$, $A_{GW}$ is the strain amplitude, and $\gamma_{GW}$ is the spectral index. NANOGrav reports a 5-frequency power law and broken power law  likelihoods in the parameter space of $(\gamma_{GW},\,A_{GW})$, as shown in the right panel of Fig.~\ref{nano}. For a possible interpretation of NANOGrav data in terms of a stochastic gravitational wave background from a stable cosmic string network,  see \cite{Ellis:2020ena,Blasi:2020mfx,Samanta:2020cdk,Blanco-Pillado:2021ygr}. In this section we discuss a possible interpretation of the NANOGrav results  in terms of a stochastic gravitational wave background from metastable cosmic strings. This possibility has recently been discussed in a supersymmetric hybrid inflation model based on a $B-L$ extension of MSSM \cite{Buchmuller:2019gfy,Buchmuller:2020lbh}. For a later more refined treatment of metastable cosmic strings see \cite{Buchmuller:2021mbb}. A non-Abelian extension of this model based on a gauge group $SU(2)_R \times U(1)_{B-L}/Z_2$ is considered in \cite{Buchmuller:2021dtt}.

Here, we briefly discuss a possible $SO(10)$ GUT embedding of the $U(1)_{B-L}$ model to realize the formation of a metastable cosmic string network.  Consider the breaking of $SO(10) \rightarrow SU(3)_c \times SU(2)_L \times U(1)_Y \times U(1)_{\chi}$ from a $45$ multiplet acquiring a non-zero vev, $v_G > v_{\chi} = M$, in suitable directions as described in \cite{Okada:2020vvb}. The charge $q_{\chi}$ associated with $U(1)_{\chi}$ is given in terms of the hypercharge $Y$ and $B-L$,
\begin{equation}
q_{\chi} = -\frac{4}{5} \, Y + B-L.
\end{equation}
Thus, the $q_{\chi}$ charges of various fields in Table~\ref{table:1} are readily obtained. Note that for the neutrino fields ($N_i,\,\mathcal{\overline N}^{c}$) and Higgs fields ($\Phi,\,\overline{\Phi}$), the $q_{\chi}$ charge coincides with $B-L$ due to their vanishing hypercharges. The Higgs fields ($\Phi,\,\overline{\Phi}$) reside
in $16+\overline{16}$ multiplets, the MSSM matter content with right-handed neutrinos in $16$-plets, and the electroweak Higgs doublet ($H_u,\,H_d$) in the $10$-plet. In SO(10) embedding, the vector-like field, $\mathcal{N}^c+\overline{\mathcal{N}}^c$, actually corresponds to a vector-like generation. The addition of a TeV scale vector-like generation in MSSM is discussed in \cite{Babu:2008ge} taking into account perturbative unification and the constraint on oblique parameters. As the superpotential $W$ in Eq.~(\ref{sup}) respects both $U(1)_Y$ and $U(1)_{B-L}$ symmetries, the $SO(10)$ embedding described here effectively leads to the $B-L$ extended MSSM after $SO(10)$ GUT breaking.

The breaking of $SO(10)$ to $SU(3)_c \times SU(2)_L \times U(1)_Y \times U(1)_{\chi}$ yields monopoles carrying MSSM as well as $U(1)_{\chi}$ charges. It's the latter ones that yield the metastable string network which decays via the Schwinger production of monopole-antimonopole pairs with a rate per string unit length of \cite{Buchmuller:2019gfy}
\begin{equation}
\Gamma_d =\frac{\mu}{2\pi} \exp(-\pi k ), \quad  k=\frac{m^2}{\mu_s},
\end{equation}
where $m \sim v_G$ is the monopole mass and $\mu_s \sim M^2$ is given by Eq.~(\ref{cosmicmu}). The parameter, $\sqrt{k} \sim v_G/M$, quantifies the hierarchy of the $SO(10)$ breaking scale $v_G$ and the $U(1)_{B-L}$ breaking scale $M$.

For the sake of completeness we briefly describe the basic ingredients and assumptions behind the numerical predictions presented in the right panel of Fig.~\ref{nano}. We employ the following expression for the gravitational wave spectrum from a cosmic string network \cite{Auclair:2019wcv}
\begin{align}
\Omega_\text{GW}(f) = \frac{8 \pi f (G \mu_s)^2}{3 H_0^2} \sum_{n = 1}^\infty \frac{2 n}{f^2} \int_{z_\text{min}}^{z_\text{max}}dz\: \, P_n \frac{\mathcal{N}\left(\ell\left(z\right),\,t\left(z\right)\right)}{H\left(z\right)(1 + z)^6} \,,
\label{eq:Omega}
\end{align}
where  $H_0 = 100 \,h\,\textrm{km}/\textrm{s}/\textrm{Mpc}$ is the present Hubble parameter, and $P_n \simeq 50/\zeta[4/3] \,n^{-4/3}$ is the power spectrum of GWs emitted by the $n^{\rm th}$ harmonic of a cosmic string loop (assuming cusps as the main source of GW emission). For the number density of cosmic string loops, $\mathcal{N}(\ell,t)$ with $\ell = 2n/((1 + z) f)$,
we use the approximate expressions of Blanco-Pillado-Olum-Shlaer (BOS) model given in \cite{Blanco-Pillado:2013qja,Blanco-Pillado:2017oxo}. For our region of interest the dominant contribution is obtained from the loops generated during the
radiation-dominated era. For $t(z)$ and $H(z)$, we use the expressions given in \cite{Blanco-Pillado:2017oxo} assuming a standard thermal history of Universe while ignoring the changes in the number of effective degrees of freedom with $z$. 
The integration range in the above equation corresponds to the lifetime of the cosmic string network, from its formation at $z_\text{max} \simeq T_r/(2.7~\text{K})$ until its decay at $z_\text{min}$  given by \cite{Leblond:2009fq,Buchmuller:2019gfy},
\begin{equation}
z_\text{min} = \left( \frac{70}{H_0}\right)^{1/2} \left( \Gamma  \; \Gamma_d  \; G \mu_s \right)^{1/4} \,,
\label{zmin}
\end{equation}
where $\Gamma \simeq 50$ and we fix the reheat temperature at $T_r=10^6$~GeV. Note that in this section $z$ refers to the red-shift and not the normalized waterfall Higgs field described in earlier sections.

The predicted range, $10^{-8} \lesssim G\mu_s \lesssim 10^{-5}$, of both cases of our model is shown in the left panel of Fig.~(\ref{nano}), where the upper bound $G\mu_s \lesssim 10^{-5}$ corresponds to the largest possible value of $r \lesssim 0.01$ in the case with $\gamma > 0$, $\delta < 0$ and the lower bound, $G\mu_s \gtrsim 10^{-8}$, is derived from the constraint $\delta \lesssim 30$. For a comparison with the NANOGrav data we first evaluate $\Omega_\text{GW}(f)$ at the first five frequencies used in the analysis of \cite{Arzoumanian:2020vkk}
and then perform a least squares power-law fit for the parameters $A_{GW}$ and $\gamma_{GW}$. The results are presented in the right panel of Fig.~\ref{nano} for a given range of $k$ and $G\mu_s$ together with the $1$-$\sigma$ (dotted) and $2$-$\sigma$(dashed) 5-frequency power law (orange) and broken power law (blue) likelihood contours from the NANOGrav 12.5-yr 5-frequency chain data \cite{Arzoumanian:2020vkk}.
Note that for $\sqrt{k} \sim 8$, a part of the predicted range, $10^{-8} \lesssim G\mu_s \lesssim 10^{-6}$,  lies within the $2$-$\sigma$ bounds of the  NANOGrav 12.5-yr 5-frequency chain data \cite{Arzoumanian:2020vkk}, as shown in the right panel of Fiq.~(\ref{nano}). The CMB constraint, $G\mu_s < 2.4 \times 10^{-7}$, is only valid for cosmic strings with lifetime exceeding CMB decoupling with $\sqrt{k} \gtrsim 8.6$. These results are consistent with the results obtained in 
\cite{Buchmuller:2020lbh}.

In summary, a large part of parameter region which is consistent with leptogenesis and  inflationary constraints lies within the 2-$\sigma$ bounds of the NANOGrav 12.5-yr data as can be deduced from the right panel of Fig.~\ref{nano}. Moreover, these results are also compatible with the recent data from pulsar timing array (PPTA) experiment  \cite{Goncharov:2021oub}. However, it is important to mention
that these results are not compatible with the bounds on $G\mu_s$
by LIGO/Virgo searches at high frequencies for a stochastic gravitational-wave background for sufficiently large $k$; see Fig.~3 in \cite{Buchmuller:2019gfy}. Assuming somewhat lower values of $\sqrt{k} \lesssim 6$, while giving up on NANOGrav evidence, our whole predicted range of $G\mu_s$ is consistent with the bounds  from the LIGO/VIRGO/KAGRA experiment \cite{LIGOScientific:2019vic,LIGOScientific:2021nrg,LIGOScientific:2021yuo}.
An interesting possibility to make the NANOGrav results compatible with the LIGO/VIRGO/KAGRA experiment \cite{LIGOScientific:2019vic,LIGOScientific:2021nrg,LIGOScientific:2021yuo} could be realized by assuming strings to re-enter the horizon at adequately late times, as is discussed in \cite{Lazarides:2021uxv}.
\section{Conclusions}
We construct a realistic model of sneutrino tribrid inflation based on a $U(1)_{B-L}$ extension of MSSM with two feasible choices $\gamma < 0$, $\delta > 0$ and  $\gamma > 0$, $\delta < 0$. For the choice $\gamma < 0$, $\delta > 0$ ($\gamma > 0$, $\delta < 0$), the predictions for the tensor to scalar ratio and the running of the scalar spectral index are given by, $3 \times 10^{-11}\lesssim r\lesssim 7\times 10^{-4}$ ($ 6 \times 10^{-7} \lesssim r \lesssim 0.009 $) and  $-0.00022 \lesssim dn_s/d\ln k \lesssim -0.0026$ ($-0.00014 \lesssim dn_s/d\ln k  \lesssim 0.005$), with the scalar spectral index, $n_s = 0.966$, as reported by Planck 2018. The corresponding ranges of model parameters are, $0.00085\lesssim \kappa \lesssim 1$ ($0.0155 \lesssim \kappa \lesssim 0.92 $),  $ 3 \times 10^{14}\lesssim M/ \text{GeV}\lesssim 5.3 \times 10^{15}$ ($ 5.9 \times 10^{15}\lesssim M/ \text{GeV}\lesssim 2 \times 10^{16}$), $ -0.005\lesssim \gamma \lesssim  -1.75$ ($0 \lesssim \gamma \lesssim 0.044$) and $ 10^{-6}\lesssim \delta \lesssim 30$ ($-2 \lesssim \delta \lesssim -0.0012$), with $y_c = 1$. These ranges are consistent with the sub-Planckian field values of the inflaton, $ 0.001\lesssim N_0/m_{P} \lesssim 1$ ($ 0.023 \lesssim N_0/m_P \lesssim 1 $). A successful realization of reheating and non-thermal leptogenesis is achieved with reheat temperature as low as $10^6 \text{ GeV}$. The range of the sneutrino inflaton mass, $ 5 \times 10^{13}\lesssim M_R^I / \text{GeV}\lesssim 5.0 \times 10^{15}$ ($ 1.3 \times 10^{14} \lesssim M_R^I / \text{GeV} \lesssim 5.5 \times 10^{15}$), gives rise to tiny neutrino masses via the seesaw mechanism. The $Z_2$ matter parity, which avoids the rapid proton decay, arises naturally as a subgroup of $U(1)_R$ symmetry. An GUT embedding of this model in $SO(10)$ is briefly described, which naturally leads to the production of metastable cosmic string network  that predicts a stochastic gravitational wave background. With string tension, $10^{-8} \lesssim  G\mu_s \lesssim 10^{-6}$, the most part of the predicted range of choice $\gamma < 0$, $\delta > 0$,  lies within the 2-$\sigma$ bounds of the recent NANOGrav 12.5-yr data. On the other hand,  a significant part of the choice $\gamma > 0$, $\delta < 0$ leads to an observable range of primordial gravitational waves from inflation with $r \gtrsim 10^{-3}$,
\cite{PRISM:2013ybg,Matsumura:2013aja,Kogut:2011xw,CORE:2016ymi}.
\section*{Acknowledgment}
This work is partially supported by the DOE grant No.~DE-SC0013880 (Q.S.). M. R. thanks Adeela Afzal for valuable discussions related to gravitational waves from the metastable string network.

\end{document}